\documentclass[twocolumn]{article}

\usepackage[modulo]{lineno}
\modulolinenumbers[1]

\usepackage[latin9]{inputenc}
\usepackage{amsmath}
\usepackage{amssymb}
\usepackage{graphicx}

\usepackage{subfigure}
\usepackage{array}
\usepackage{natbib}
\usepackage[thinspace,thinqspace,squaren,textstyle]{SIunits}
\usepackage{widetext}

\makeatletter\@ifundefined{date}{}{\date{}}
\makeatother

\evensidemargin\oddsidemargin \hoffset-22.4mm \voffset-28.4mm
\begin{document}

\title{On the Dynamics of the Flow and the Sound Field of an Organ Pipes's Mouth Region}

\author{Jost Leonhardt Fischer$^{1)}$, Rolf Bader$^{1)}$, Markus Abel$^{2),3)}$ \\
$^{1)}$ Institute of Systematic Musicology, University Hamburg, Neue Rabenstr. 13 \\ D-20354 Hamburg, Germany. jost.leonhardt.fischer@uni-hamburg.de\\
$^{2)}$ Department of Physics and Astronomy, Potsdam University, Karl-Liebknecht-Str. 24\\ D-14476, Potsdam-Golm, Germany.\\
$^3$Ambrosys GmbH, Potsdam, Germany.}

\maketitle\thispagestyle{empty}

\begin{abstract}
The dynamics of an organ pipe's mouth region has been studied by numerical simulations. The investigations presented here were carried out by solving the compressible Navier-Stokes equations under suitable initial and boundary conditions using parts of the open source C++ toolbox OpenFOAM. The focus of the study is on the examination of the velocity field close to the jet. The components of the velocity field sampled at a cross-section spanning the mouth were analyzed by using methods of coarse-graining. It is shown that the dynamics of the sampled velocity field can be separated into two fractions, a velocity component mainly comprising the jet's flow velocity and a component that essentially carries particle velocity caused by back-propagating sound waves inside the resonator. The SPL-spectra calculated of the data sets of the numerical simulations are compared with the measurements on real organ pipes. The phase-space representations of sound pressure and particle velocity indicate the presence of nonlinearities in the organ pipes's mouth region. The numerical results are consistent with measurements on real organ pipes.
\end{abstract}

\section{Introduction}
\label{sec:indroduction}
 \begin{figure*} [!htb]
\centering
\subfigure[]
{
\includegraphics[draft=false,width=0.12\textwidth]{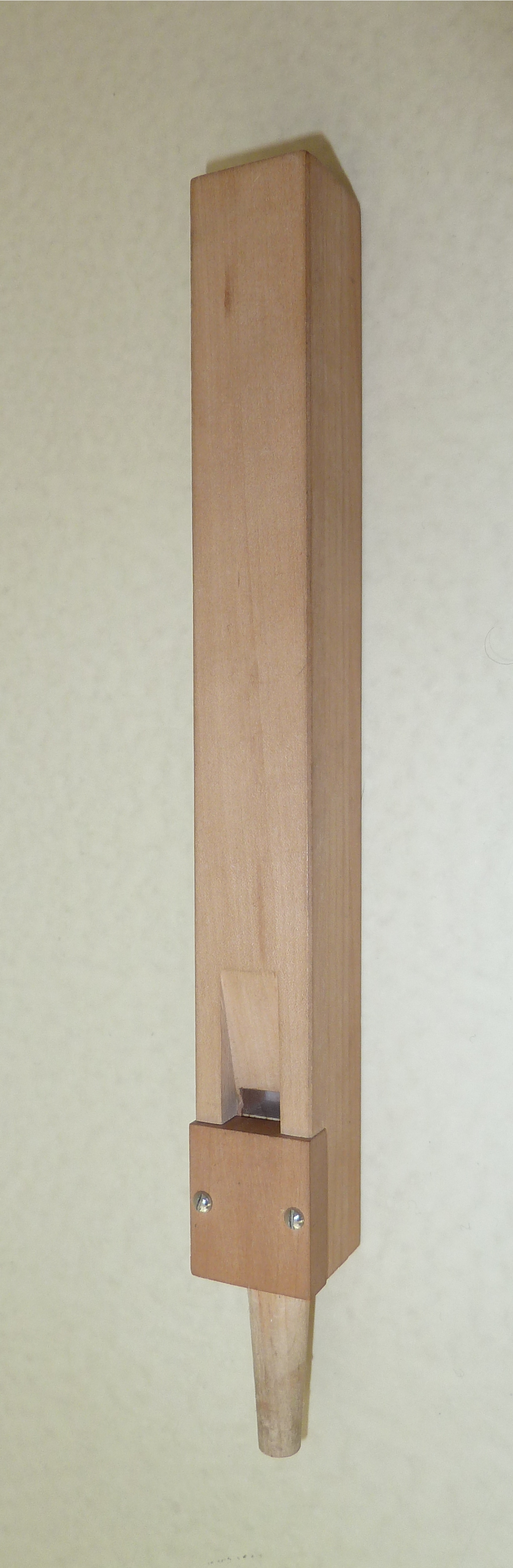}
\label{fig:numSim01}
}
\subfigure[]
{
\includegraphics[draft=false,width=0.60\textwidth]{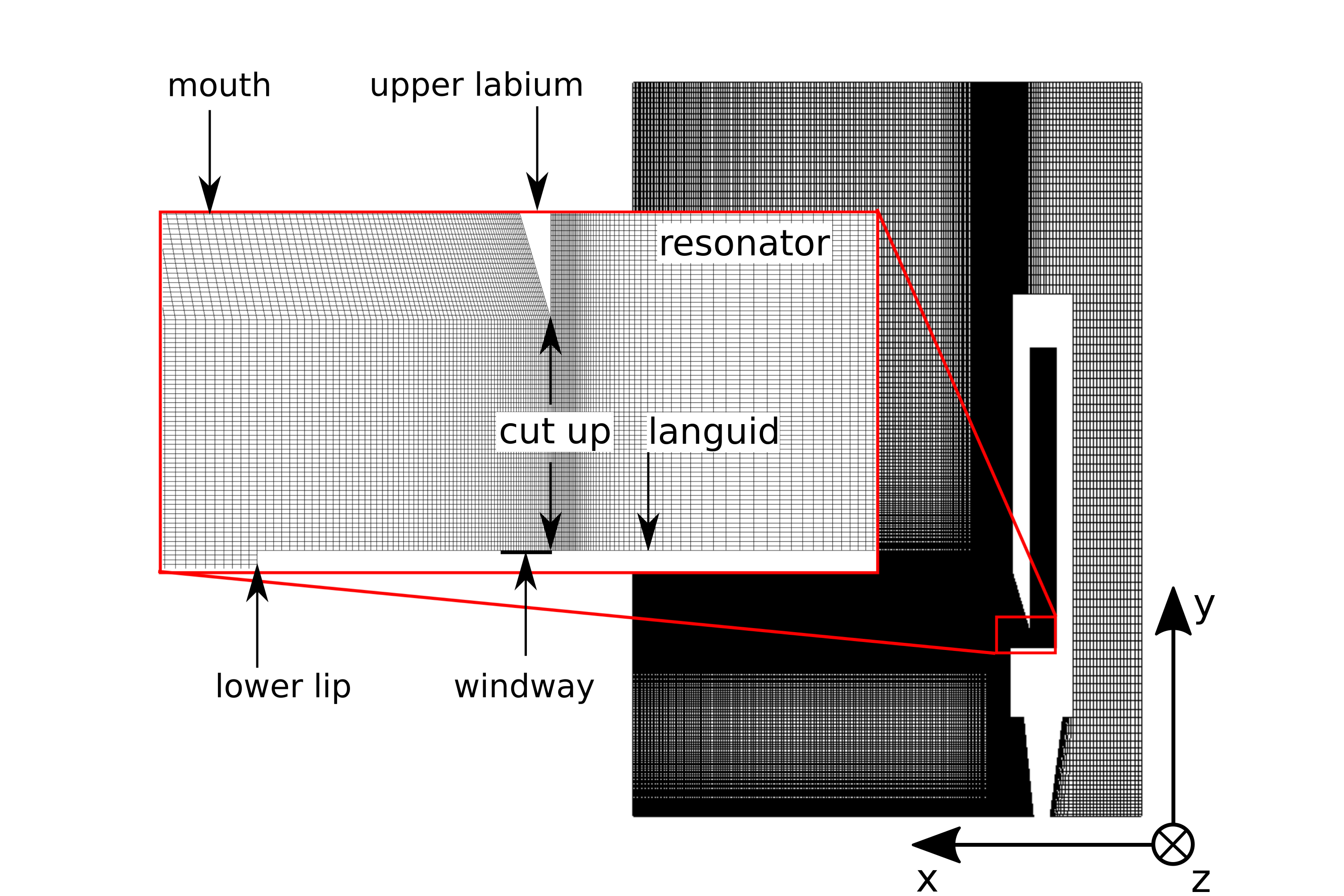}
\label{fig:numSim02}
}
\caption{(a) A stopped wooden organ pipe provided by German organ builder Alexander Schuke Orgelbau GmbH. The dimensions of the resonator (length $\times$ width $\times$ depth) are $(120\usk\milli\metre \times 9\usk\milli\metre \times 9\usk\milli\metre$). The width of the windway is $0.6\usk\milli\metre$. The length of the cut-up is $5.5\usk\milli\metre$. The pipe geometry is used as a template for the numerical set-up. (b) The computational grid of the organ pipe and its surrounding. The detail shows the refinement and the grading of the mesh in the mouth region. The smallest cells with $\delta x \leq 0.1\usk\milli\metre$ are in the cut-up.  }
\end{figure*}
The mouth region of an organ pipe is the domain where the oscillating air sheet, called the jet, acts. The mouth is enclosed by the upper labium, the lower lip, the languid and the lower resonator region, cf. Fig.\,\ref{fig:numSim01},~\ref{fig:numSim02}. In the mouth region interactions between the jet's flow field and the sound field take place, which constitute the sound generation of the instrument. Due to the fact that the jet of an organ pipe is a spatially extended, turbulent coherent structure, the dynamics of the region of interest is very difficult to determine experimentally. 
Flow field visualization of the initial transient in a small recorderlike flue organ pipe has been studied by Verge et al.\cite{Verge_1994}. Experimental investigations of sound production in recorderlike instruments have been done by Fabre et al.\cite{Fabre-97} 
A general description of the mechanics of fluids, the fluid dynamics and the aeroacoustics of an organ pipe's jet in particular, is given by the compressible Navier-Stokes equations\cite{schlichting2003, morse1968theoretical}, hereafter referred to as CNSE, with suitable boundary and initial conditions. Several numerical investigations regarding wind instruments were performed in recent years. Miyamoto, Ito, and Takahashi\cite{miyamoto2010applicability} investigated sound vibration of an air-reed instrument by solving the CNSE numerically using a Large-Eddy-Simulation model (LES).
Bader\cite{bader2013nonlinearities} investigated flute-like instruments by numerical simulations using a Reynolds Averaged Navier-Stokes finite element model (RANS-FEM). 

The focus of the present work is on reproducing the complex behavior of the jet and other turbulent coherent structures in the mouth region of an externally non-driven organ pipe. The aim is to characterize the dynamics with respect to the velocity components of the flow field and the sound field close to the jet. With methods of coarse-graining, principle aspects of the dynamics are analyzed to get a simpler picture of the complex dynamics in the mouth region of an organ pipe.

\section{Implementation, Run and Post-Processing}
\label{sec:implementation}

The numerical treatment of compressible problems is still an advanced task. The main difficulties arise when reproducing the interactions between the flow field and the sound field~\cite{Howe-75, Fabre-Hirschberg-96, Fabre-00}. Numerical simulations allow to study the dynamics of the jet and the resonator simultaneously. In general, a successful procedure of realization can be divided into the following sections: physical previews, cf. Eqs.\,(\ref{eqn:continuity_equation})--(\ref{eqn:isentropic_coefficient}), pre-processing, cf. Eqs.\,(\ref{eqn:Reynolds})--(\ref{eqn:model_constant_C_k}), Tab.\,\ref{tab:thermophys_Eigenschaften}, processing and post-processing~\cite{fischer2014nichtlineare}. 

The present numerical simulations were realized by using parts of the C++ toolbox OpenFoam-3.0.0~\cite{openfoam_guide}. These libraries include customized numerical solvers as well as pre- and post-processing utilities for the solution of continuum mechanics problems, including computational fluid dynamics (CFD) and computational aeroacoustics (CAA). The code is released as free and open source software under the GNU General Public License. General aspects about pre-processing, run and post-processing are documented in the OpenFOAM User Guide as well as in the OpenFOAM Programmer Guide~\cite{openfoam_guide}.

The geometry of a stopped wooden organ pipe, produced and provided by the German organ builder Alexander Schuke Orgelbau GmbH\cite{Schuke}, is transferred into a structured 2D computational grid, cf. Figs.\,\ref{fig:numSim01},~\ref{fig:numSim02}. 
The organ pipe's surfaces are considered as acoustically inert (boundary condition: no slip). The outer limitations of the mesh are configured as open. This means that the radiated sound as well as all other physical quantities can propagate through the boundaries without any restrictions. The mesh size (length~$\times$~width~$\times$~depth) is ($260\usk\milli\metre\times180\usk\milli\metre\times1\usk\milli\metre$) with $254342$ mesh points, $505170$ faces and $126000$ hexahedra. The technique how to write a proper mesh file and how to generate a mesh can be found in the OpenFOAM User Guide~\cite{openfoam_guide}.

\begin{widetext}
\begin{align}
\frac{\partial \rho}{\partial t} + \nabla \cdot \left( \rho \boldsymbol{u} \right) = 0 &&\text{(continuity equation)}\label{eqn:continuity_equation}&\\
\rho \left(\frac{\partial \boldsymbol{u}}{\partial t} + \left( \boldsymbol{u} \cdot \nabla \right) \boldsymbol{u} \right) + \boldsymbol{u}\frac{\partial \rho}{ \partial t} = -\nabla p + \left( \lambda + \eta  \right) \nabla \left( \nabla \cdot \boldsymbol{u}  \right) + \eta \mathrm{\Delta} \boldsymbol{u}+\rho \cdot \boldsymbol{g} && \text{(momentum balance, CNSE)}\label{eqn:momentum_balance}&\\
\frac{\partial (\rho E)}{\partial t} + \nabla \cdot \left(\rho E \boldsymbol{u} \right) = - \nabla \left( p \boldsymbol{u} \right) + \nabla \left( \boldsymbol{\tau} \boldsymbol{u}  \right) - \nabla\boldsymbol{q} &&\text{(energy balance)}\label{eqn:energy_balance}&\\ 
\text{with}\qquad \boldsymbol{q} = - \kappa \nabla T &&\text{(Fourier's law, $\kappa$: thermal conductivity)}\label{eqn:Fouriers_Law}&\\
E  = e + \frac{1}{2} |\boldsymbol{u}|^2 &&\text{(energy)}\label{eqn:energy}&\\
H  =e+\frac{p}{\rho} = C_pT && \text{(calorimetric state equation, enthalpy)}\label{eqn:enthalpy}&\\
p  =(\gamma - 1) \rho e && \text{(closure)}\label{eqn:pressure_closure}&\\
\gamma = \frac{C_p}{C_V}\approx1.4 && \text{(isentropic coefficient)}\label{eqn:isentropic_coefficient}&
\end{align}\\
\end{widetext}

\begin{table} [!htb]
\begin{center}
\begin{tabular}{|  l  |  l  |  c  |}
\hline
Property & Value & Unit\\\hline 
  Molecules & $1$ & \\ 
  Molar mass $M$ & $28.9$ & $\usk\gram\per\mole$\\
  Specific heat capacity & &\\(p=\textit{const}) $C_p$ & $1007$ & $\usk\joule\per(\kilo\gram \cdot \kelvin)$\\
  Dynamic viscosity $\mu$  & $1.8\cdot 10^{-5}$ & $\usk\pascal \cdot \second$\\
  Prandtl number $Pr$ & $0.72$ & - \\\hline 
\end{tabular}
\end{center}
\caption{Implemented thermo-physical properties.}
\label{tab:thermophys_Eigenschaften}
\end{table}

The initial conditions ($t=t_0$) for the physical properties to be calculated are implemented, e.g. the velocity profile of the flow field at the windway, which is assumed to be uniform (hat-profile), cf. Fig.\,\ref{fig:numSim02}: $\boldsymbol{u}(x,y,z,t_0)=18\usk\metre\per\second \cdot \boldsymbol{\mathbf{e_y}}=u_y$, the value of the initial pressure field $p(x,z,y,t_0)=1013.25\usk\hecto\pascal$, the value of the initial temperature field $T(x,y,z,t_0)=293\usk\kelvin$ etc. Additional thermo-physical properties have to be taken into account, e.\,g. the molar mass $M$ of the medium air, which is assumed as a perfect gas, the heat capacity at constant pressure $C_p$, the dynamic viscosity $\mu$ and the Prandtl number $Pr$, which will be discussed subsequently. The chosen values of the thermo-physical properties of the medium air are summarized in Tab.\,\ref{tab:thermophys_Eigenschaften}. 

A necessary step is to estimate the relevant characteristic fluid mechanical numbers of the given set-up to fit the numerical implementation of the problem.

The probably most important fluid mechanical characteristic number is the Reynolds-number. The Reynolds-number is the dimensionless ratio of inertial forces to viscous forces\cite{schlichting2003}. The forces are represented by the characteristic length $L$ and the characteristic velocity $U$ of the flow field on the one hand and the kinematic viscosity on the other, cf. Eq.\,(\ref{eqn:Reynolds}). The Reynolds-number is used to valuate the transition of laminar to turbulent in a particular flow field, e.g. in a pipe flow with diameter $d$ turbulent flow occurs if $Re_d>4000$~\cite{schlichting2003}. The characteristic length of the problem discussed here is assumed to be the length of the cut-up, $L=5.5 \cdot 10^{-3}\usk\metre$, regarding to the free propagation length of the jet. Coherent fluid dynamical objects like vortices expected to occur in the mouth region are formed by the constraints given by the cut-up length. The characteristic velocity is assumed as $U=u_y$, the initial value of the velocity profile of the jet at the windway. The kinematic viscosity $\nu=\mu/\rho$ of the medium air is about $\nu~=~1.53 \cdot 10^{-5}\usk\metre^2\per\second$. With these values the Reynolds-number is estimated at 

\begin{equation}
Re=\frac{L \cdot U}{\nu}= 6470
\label{eqn:Reynolds}
\end{equation}

The gained Reynolds number for the given problem indicates that the flow field in the mouth region of the organ pipe is of weak turbulence\cite{schlichting2003}. \\

As a further important characteristic number the Strouhal-number is discussed. The Strouhal-number is the dimensionless ratio between an aeroacoustical quantity, the so-called vortex shedding frequency $f$, and the fluid mechanical quantities, the characteristic length $L$ and the characteristic velocity $U$ of the flow\cite{schlichting2003}. The vortex shedding frequency in the mouth region is caused by the oscillations of the jet. Here the hypothesis is that the vortex shedding frequency corresponds to the fundamental frequency of the real operating organ pipe $f=700\usk\hertz$, which one obtains by measurements, cf. Fig.\,\ref{fig:spl_p_cs0}. The Strouhal-number is estimated at

\begin{equation}
St=\frac{L \cdot f}{U}= 0.21
\label{eqn:Strouhal}
\end{equation}

The attained Strouhal-number corresponds to values one would expect for the problem considered. \\

The Prandtl-number, already mentioned above, is defined as the ratio of kinematic viscosity, also known as momentum diffusivity, to thermal diffusivity, cf. Eq.\,(\ref{eqn:Prandtl}). The thermal diffusivity of the medium air at normal conditions is given by $\alpha=\kappa/(\rho \cdot C_p)$.  Hereby $\kappa=0.0257\usk\watt\per(m \cdot K)$ is the thermal conductivity, cf. Eq.\,(\ref{eqn:Fouriers_Law}) and $C_p$ the specific heat capacity at constant pressure conditions, cf. Eq.\,(\ref{eqn:enthalpy}). The thermal diffusivity is calculated as $\alpha=1.9 \cdot 10^{-5}\usk\metre^2\per\second$. This leads to the estimation of the Prandtl-number at

\begin{equation}
Pr=\frac{\nu}{\alpha}= 0.72
\label{eqn:Prandtl}
\end{equation}

The second essential step is to estimate the Kolmogorov microscales to evaluate the chosen computational grid sizes as well as the resolution of the numerical time step size. The Kolmogorov length scale $\eta$ and time scale $\tau_{\eta}$ are the microscales, where the viscosity dominates and the turbulent kinetic energy is dissipated into heat. The scales are defined as\cite{kolmogorov1941a}

\begin{equation}
\eta=\left(\frac{\nu^3}{\epsilon}\right)^{1/4},  \qquad \tau_{\eta}=\left(\frac{\nu}{\epsilon}\right)^{1/2}  
\label{eqn:Kolmogorov_length}
\end{equation}

where $ \epsilon=U^3 / L$ is the average rate of dissipation of turbulence kinetic energy $k$ per unit mass.  The Kolomogorov microscales are estimated at $\eta=7.62\cdot10^{-6}\usk\metre$ and $\tau_{\eta}=3.8\cdot10^{-6}\usk\second$. The time increment of the numerical simulations is $\delta t=10^{-7}\usk\second$ for reasons of numerical stability and therefore smaller   by the factor $38$ than the Kolomogorov time scale requires. Note that numerical stability of explicit time integration schemes, which are, inter alia, utilized in the numerical simulations, is given by the Courant-Friedrichs-Lewi condition (CFL-condition)

\begin{equation}
\mathrm{CFL}=\frac{|u|\cdot \delta t}{\delta x}\leq C_{max}=1
\label{eqn:Courant-Friedrichs-Lewy_condition}
\end{equation}

where $|u|$ is the estimated absolute value of the maximal velocity occurring in the numerical simulation, here the speed of sound estimated at $c_{0,max}=350\usk\metre\per\second$. The smallest grid sizes of the created mesh are $\delta x = 1 \cdot 10^{-4}\usk\metre$ and $\delta y = 2 \cdot 10^{-4}\usk\metre$. Hence $\mathrm{CFL_{max}}=0.35\leq1$ and therefore Eq.\,(\ref{eqn:Courant-Friedrichs-Lewy_condition}) is satisfied by choosing $\delta t$ as mentioned.

Compared with the Kolmogorov length scale the smallest grid sizes are too large by the factor $14$. That means, that turbulent structures smaller than the grid size cannot be resolved by the mesh. A refinement of the mesh would lead to a considerably increase of computing effort. Therefore a turbulence model is utilized to model the transport of turbulent kinetic energy $k$ into the sub-grid scales (SGS) the mesh doesn't resolve. The turbulent kinetic energy $k$ can be split up into a grid-scale term $k_{GS}$ and a subgrid-scale term $k_{SGS}$ as follows 

\begin{equation}
k= \frac{1}{2} \overline{u_k u_k} =  \underbrace{ \frac{1}{2} \overline{u}_k\overline{u}_k }_{k_{GS}} + \underbrace{ \frac{1}{2} ( \overline{u_k u_k} - \overline{u}_k \overline{u}_k )}_{k_{SGS}}
\label{eqn:turb_kin_energy_gs_sgs}
\end{equation}

using Einstein's summation convention for the spatial indices~$_k=1,2,3$ of the velocity components $u_k$. As a suitable LES-Model for the SGS turbulent kinetic energy $k_{SGS}$ (SGS-k model) a one-equation dynamic subgrid-scale model is chosen~\cite{openfoam_guide}. The model equation is by Eq.\,(\ref{eqn:model_equation_turb_kin_energy}). Further explanations of the addressed terms are given by Eqs.\,(\ref{eqn:turb_kin_energy_subgrid_scales})--(\ref{eqn:model_constant_C_k}). More detailed information about the turbulence model used can be found in the OpenFOAM User Guide as well as in the OpenFOAM Programmer Guide~\cite{openfoam_guide}.

The numerical simulations were calculated in parallel on the HPC-Cluster at the University of Hamburg using 16 nodes with a total of 256 CPUs. The OpenFOAM solver \texttt{rhoPimpleFoam} for compressible problems was used. An amount of data of about $81\usk$GB per simulation run was generated. 
The run time of one simulation was $5.5\usk\hour$. The simulation times were up to $t_s=75\usk\milli\second$. \\

Visualizations of the numerical simulations were generated using the open source multi-platform data analysis and visualization application ParaView-4.1~\cite{openfoam_guide}. 

\begin{figure*} [p]
\subfigure[]
{
\includegraphics[draft=false,width=0.31\textwidth]{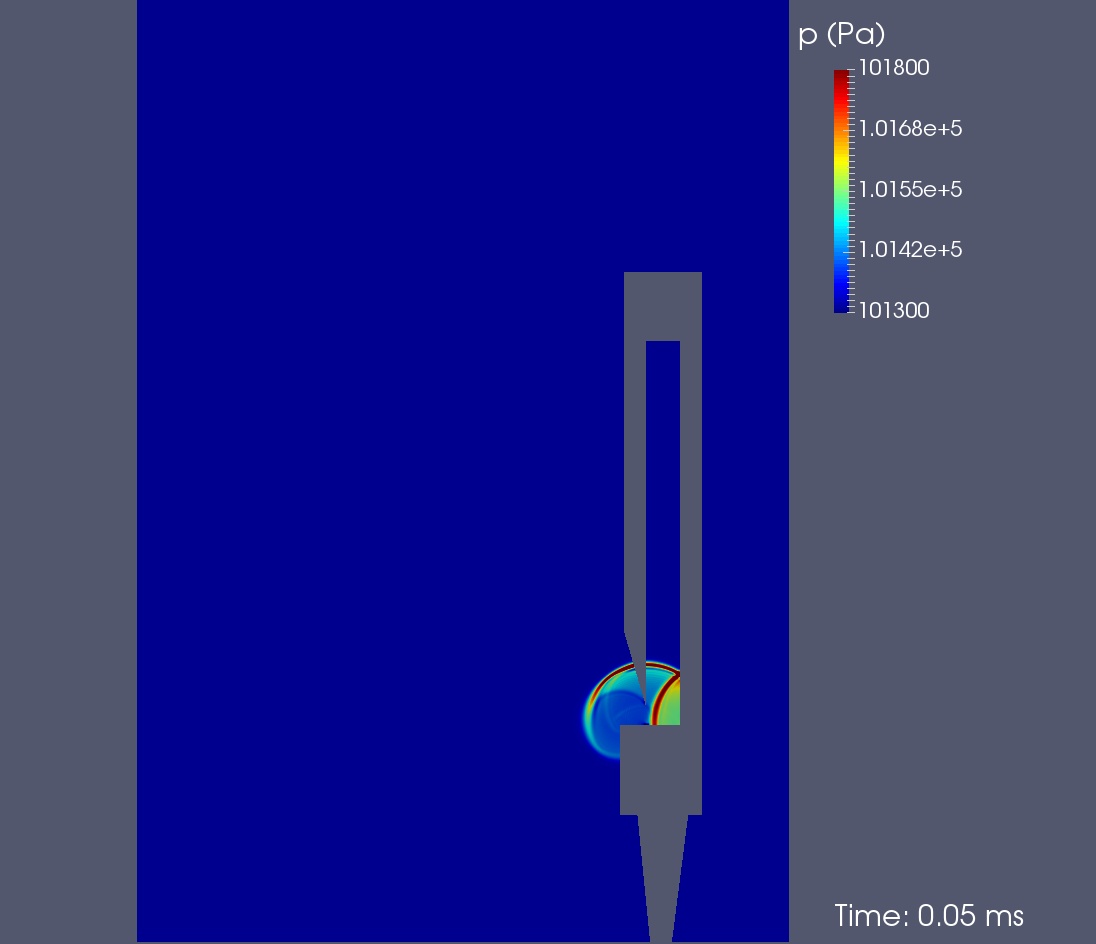}
 \label{fig:init_pressure02}
}
\subfigure[]
{ 
\includegraphics[draft=false,width=0.31\textwidth]{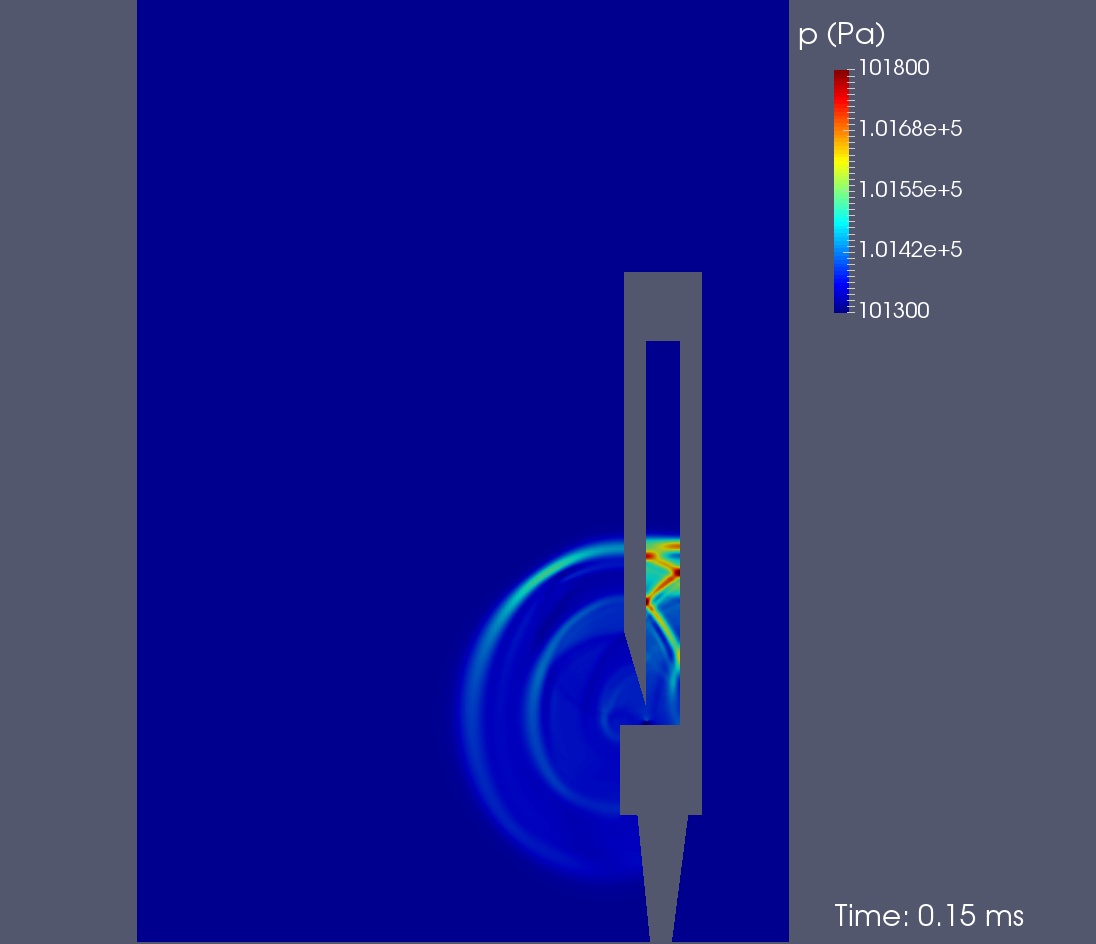}
\label{fig:init_pressure04}
}

\subfigure[]
{
\includegraphics[draft=false,width=0.31\textwidth]{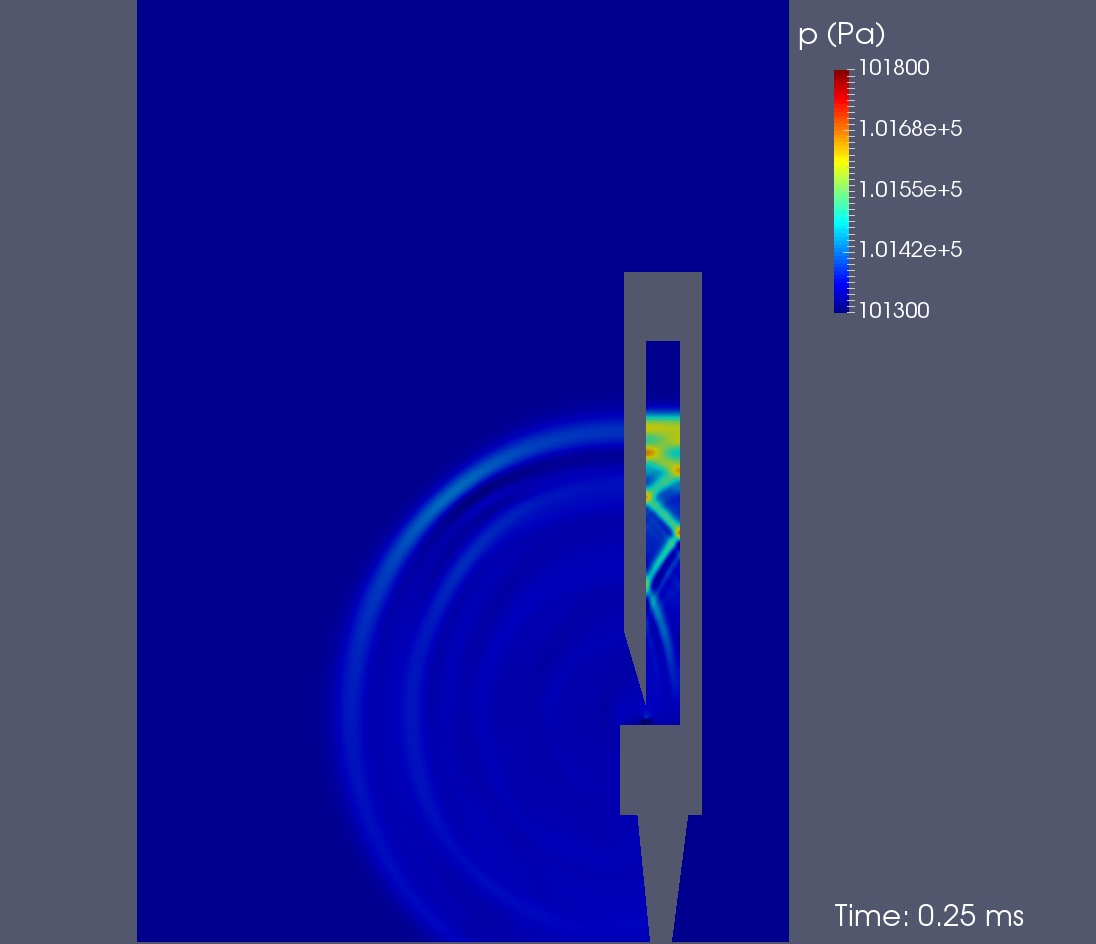}
 \label{fig:init_pressure06}
}
\subfigure[]
{
\includegraphics[draft=false,width=0.31\textwidth]{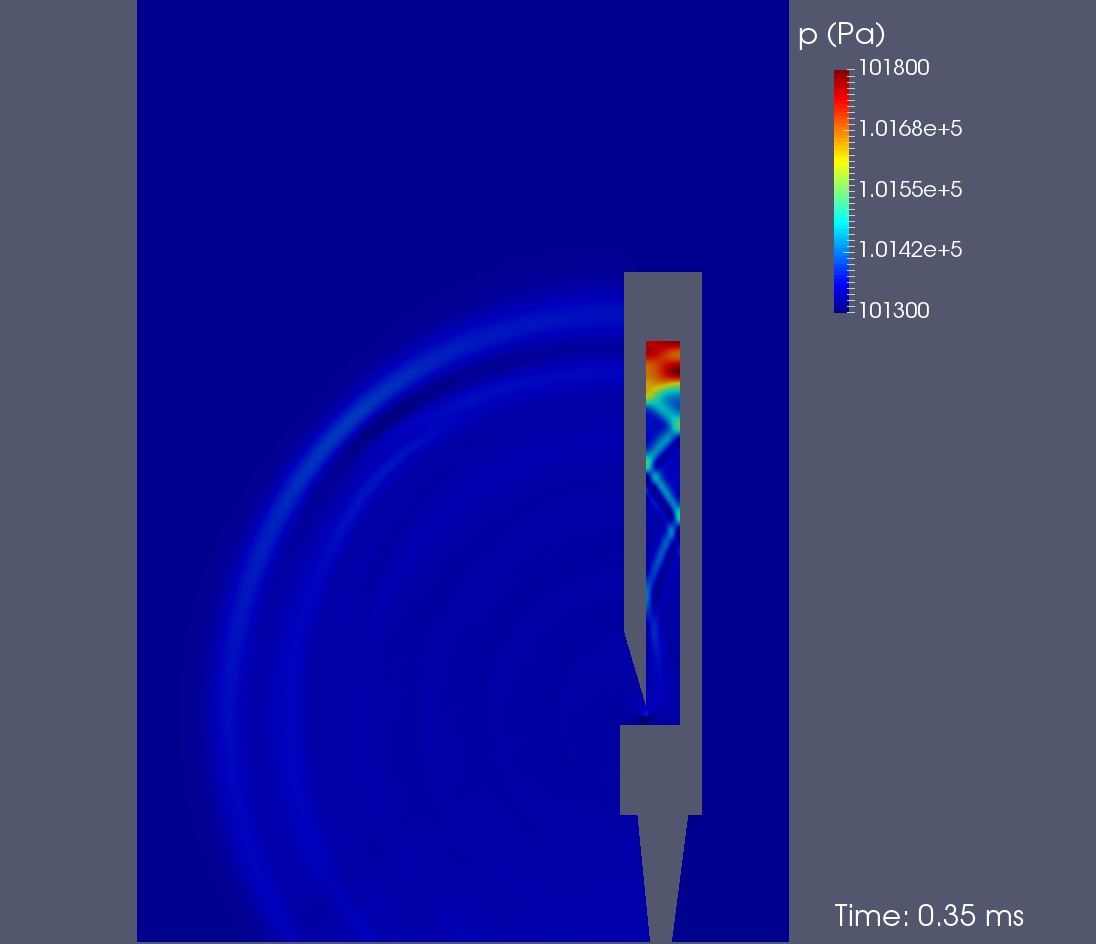}
 \label{fig:init_pressure08}
}

\subfigure[]
{
\includegraphics[draft=false,width=0.31\textwidth]{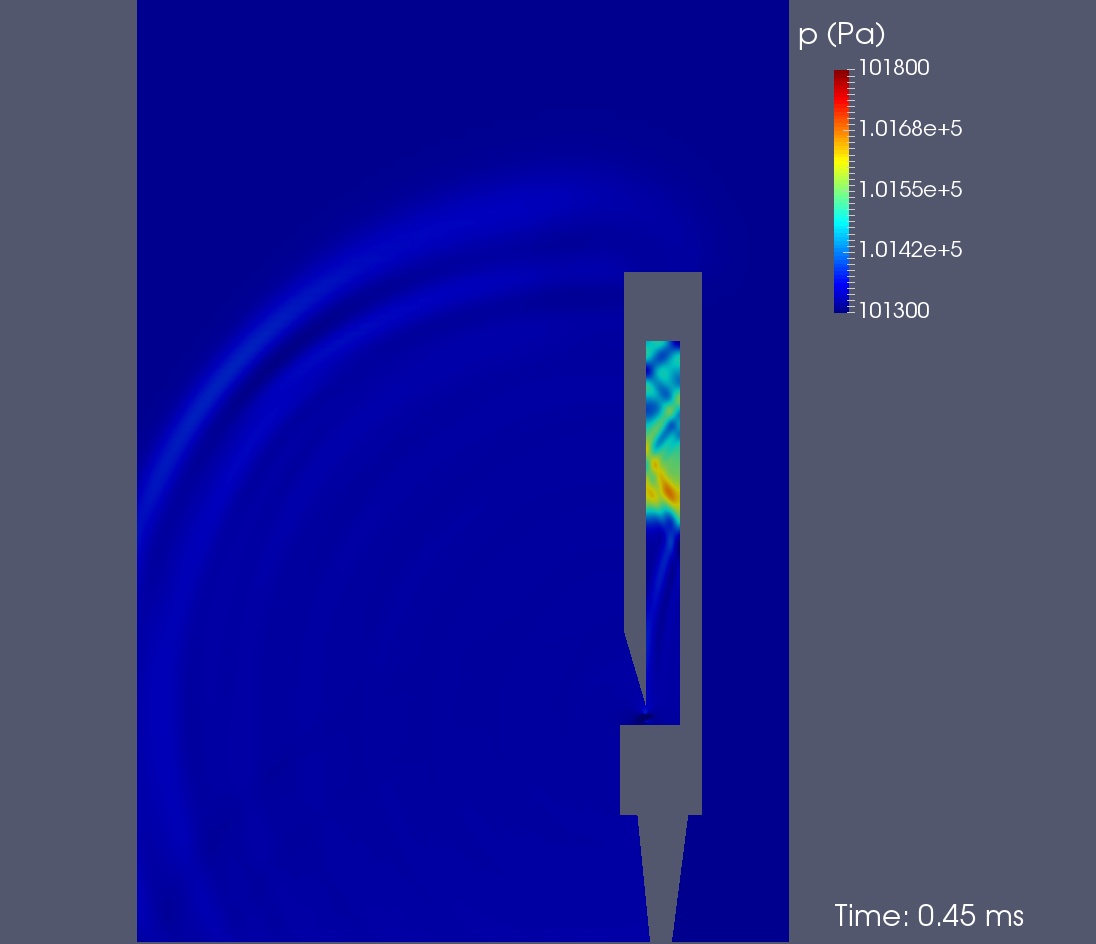}
 \label{fig:init_pressure10}
}
\subfigure[]
{
\includegraphics[draft=false,width=0.31\textwidth]{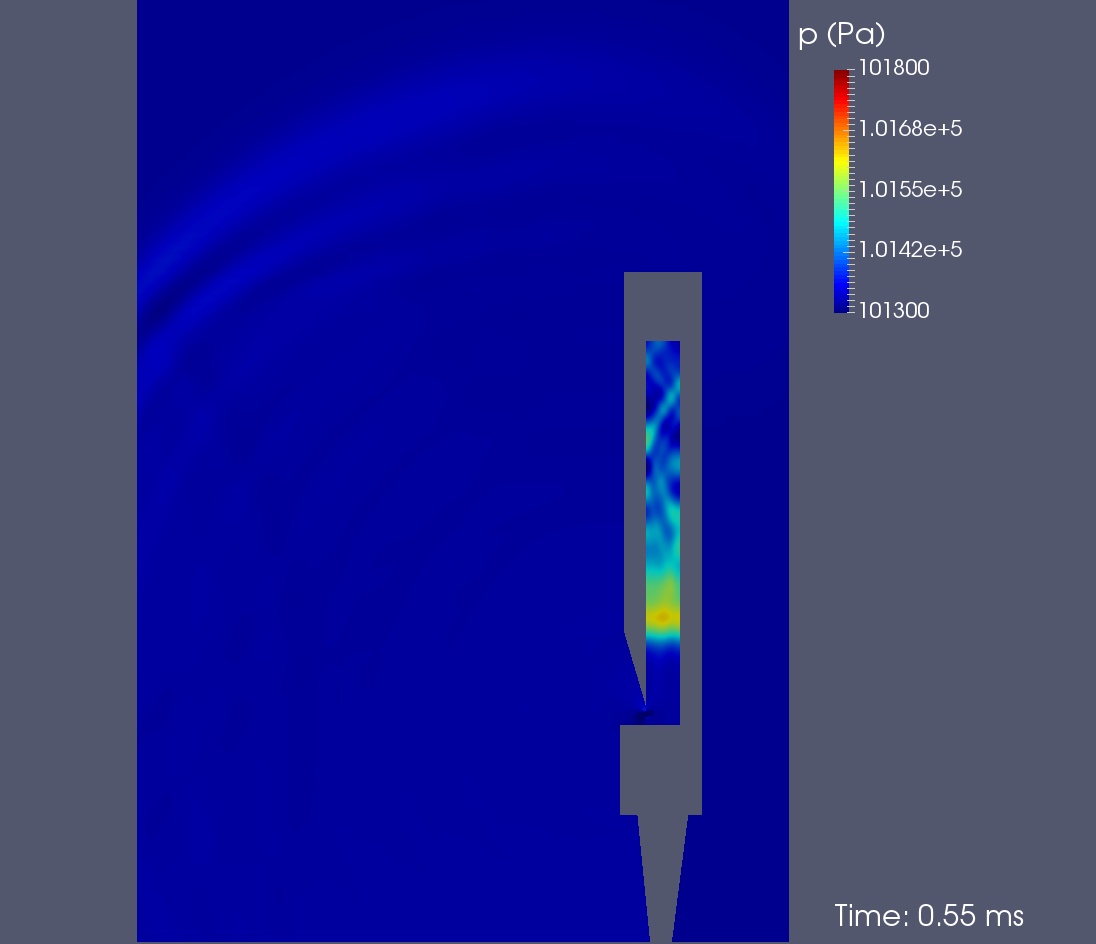}
 \label{fig:init_pressure12}
}

\subfigure[]
{
\includegraphics[draft=false,width=0.31\textwidth]{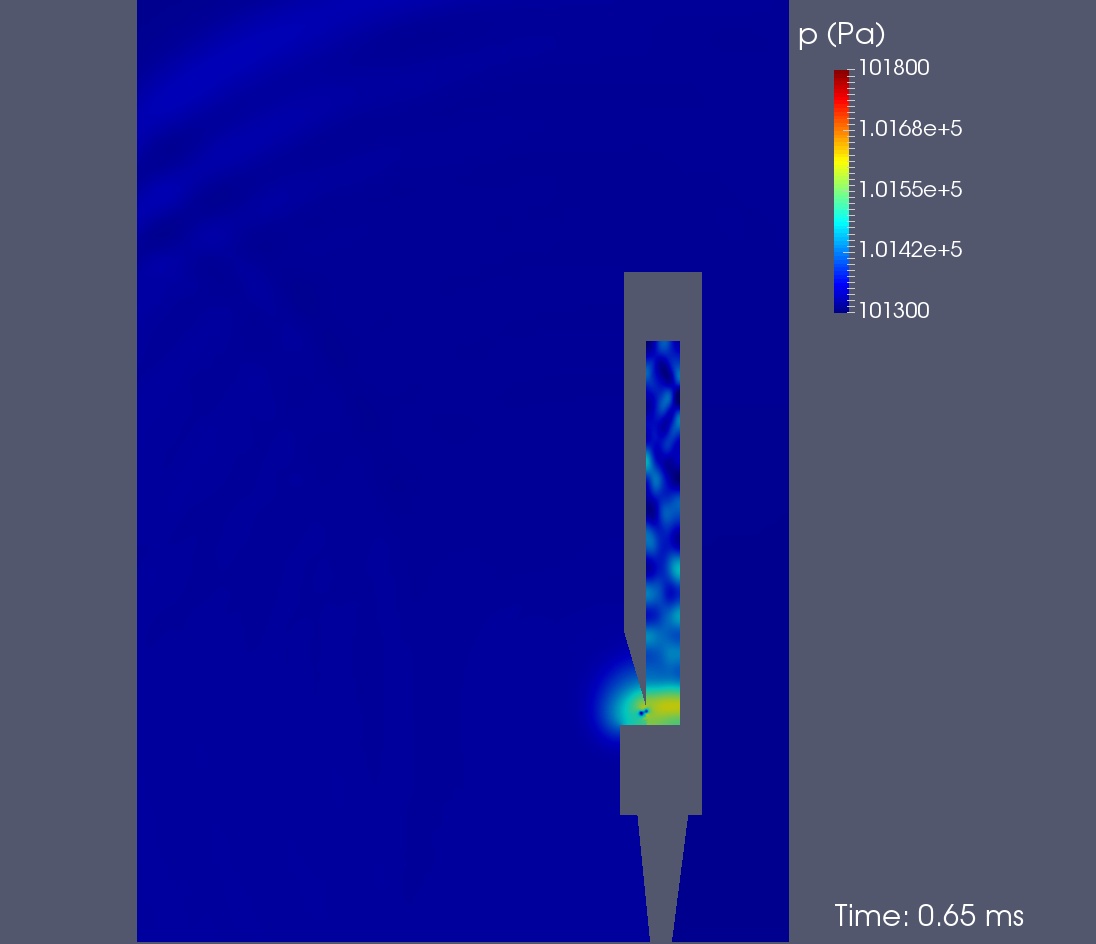}
 \label{fig:init_pressure13}
}
\subfigure[]
{
\includegraphics[draft=false,width=0.31\textwidth]{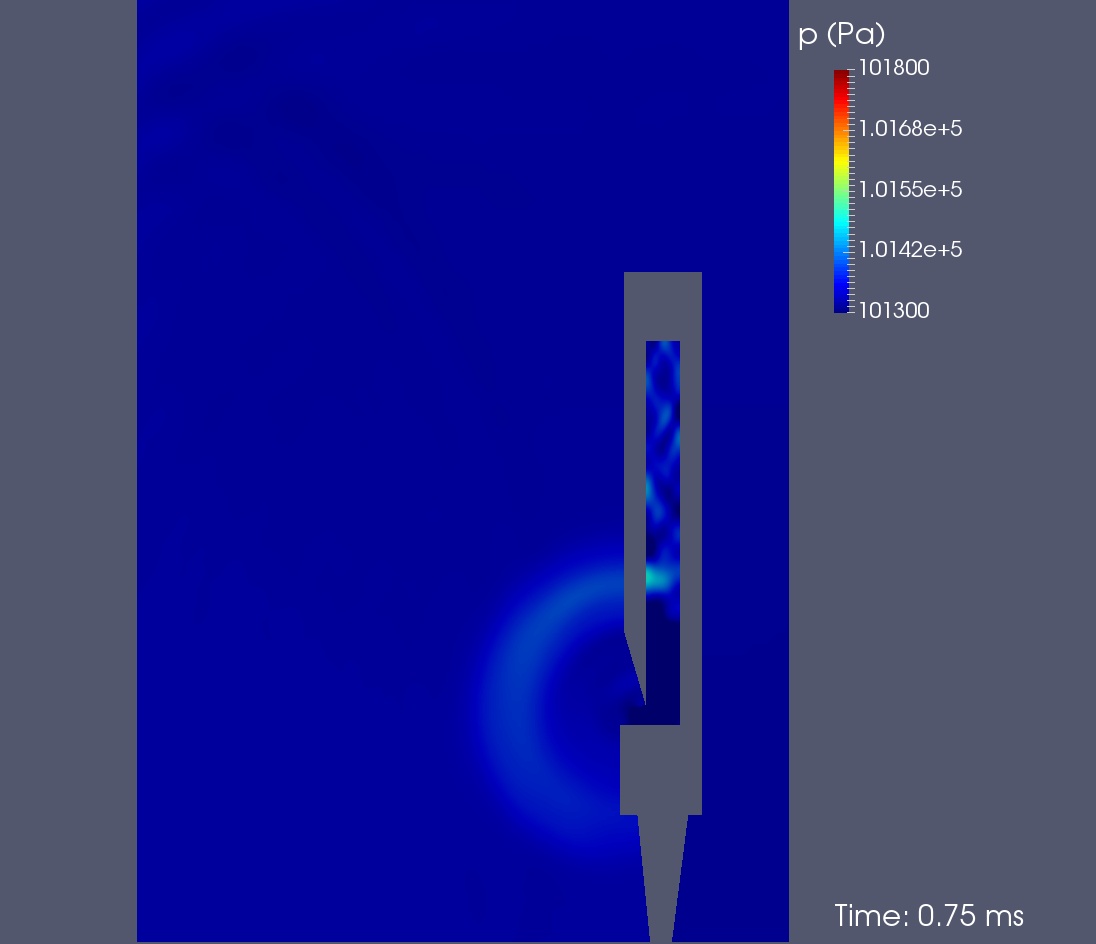}
 \label{fig:init_pressure15}
}

\caption{(a)--(h) Sequence of a numerical simulation of an operating stopped wooden organ pipe. Color-coded is the pressure $p$ in the range of $1013\usk\hecto\pascal-1018\usk\hecto\pascal$. Depicted are snapshots of the initial excitation process, $t_s=0.05\usk\milli\second-0.75\usk\milli\second$. The initial sound wave, formed by the transverse modes and damping processes at the boundary layers and in the air column, propagates with speed of sound of $c_0=333\usk\metre\per\second$ in the resonator. At the upper end the sound wave reflects and travels back to the mouth region. The sound wave radiates spherically into the free space (circular in the 2D set up).}
\label{fig:init_p}
\end{figure*}

Sequences of the initial excitation process, $t_s=0.05\usk\milli\second-0.75\usk\milli\second$, are depicted in Figs.\,\ref{fig:init_pressure02}--\ref{fig:init_pressure15}. Color-coded is the pressure $p$ in the range of $1013\usk\hecto\pascal-1018\usk\hecto\pascal$.
One can see how the initial pressure fluctuation propagates in the resonator. One observes the typical initial transverse modes which reflect at the inner walls of the resonator. Damping processes at the boundary layers and in the air column are responsible for forming a plane sound wave front (yellow, red) out of the initial transverse modes. The sound wave front propagates and gets reflected at the upper end and travels back to the mouth region. Finally the sound wave radiates as a spherical wave front into the free space (circular wave front in the 2D set-up). Clearly seen is the excitation of the pressure's amplitude inside the resonator (yellow, red and dark blue respectively) and its attenuation in the free space. The speed of sound in the numerical simulation is calculated by the ratio of the propagation length back and forth in the resonator to the duration time $c_0=0.2\usk\metre/(6\cdot10^{-4}\usk\second)=333\usk\metre\per\second$.

\begin{widetext}
\begin{equation}
\frac{\partial (\rho k_{SGS})}{\partial t} + \frac{\partial(\rho \overline{u}_j k_{SGS})}{\partial x_j} - \frac{\partial}{\partial x_j}\left[ \rho ( \nu + \nu_{SGS}) \frac{\partial k_{SGS}}{\partial x_j} \right] = -\rho \tau_{ij} : \overline{D}_{ij} - C_{\epsilon} \frac{\rho k_{SGS}^{3/2}}{\Delta}
\label{eqn:model_equation_turb_kin_energy}
\end{equation}

with
\begin{align}
k_{SGS} = \frac{1}{2} \tau_{kk} =\frac{1}{2}\left(  \overline{u_k u_k} - \overline{u}_k \overline{u}_k\right) && \text{(turb. kin. energy of the subgrid scales)}\label{eqn:turb_kin_energy_subgrid_scales}&\\
-\rho \tau_{ij} : \overline{D}_{ij} = -\frac{2}{3} \rho k_{SGS} \frac{\partial \overline{v}_k}{\partial x_k}  + \rho \nu_{SGS} \frac{\partial \overline{u}_i}{\partial x_j} \left( 2 \overline{D}_{ij} - \frac{1}{3} tr \left( 2 \overline{D}\right)\,\delta_{ij} \right) && \text{(production term)} \label{eqn:production_term}&\\
\overline{D}_{ij} = \frac{1}{2} \left( \frac{\partial \overline{u}_i}{\partial x_j} + \frac{\partial \overline{u}_j}{\partial x_i} \right) && \text{(resolved-scale strain rate tensor)}\label{eqn:resolved-scale_strain_rate_tensor}&\\
 C_{\epsilon}=1.05 &&\text{(model constant)}\label{eqn:model_constant_C_epsilon}&\\
\Delta \quad &&\text{(Sauter mean diameter of the grid cell)}\label{eqn:Sauter_mean_diameter}&\\
\nu_{SGS} = C_k k_{SGS}^{1/2}\Delta &&\text{(subgrid scale eddy viscosity)}\label{eqn:SGS_eddy_viscosity}&\\
 C_k = 0.07 && \text{(model constant)}\label{eqn:model_constant_C_k}&\\\nonumber
\end{align}
\end{widetext}

\begin{figure*} [p]
\subfigure[]
{
\includegraphics[draft=false,width=0.31\textwidth]{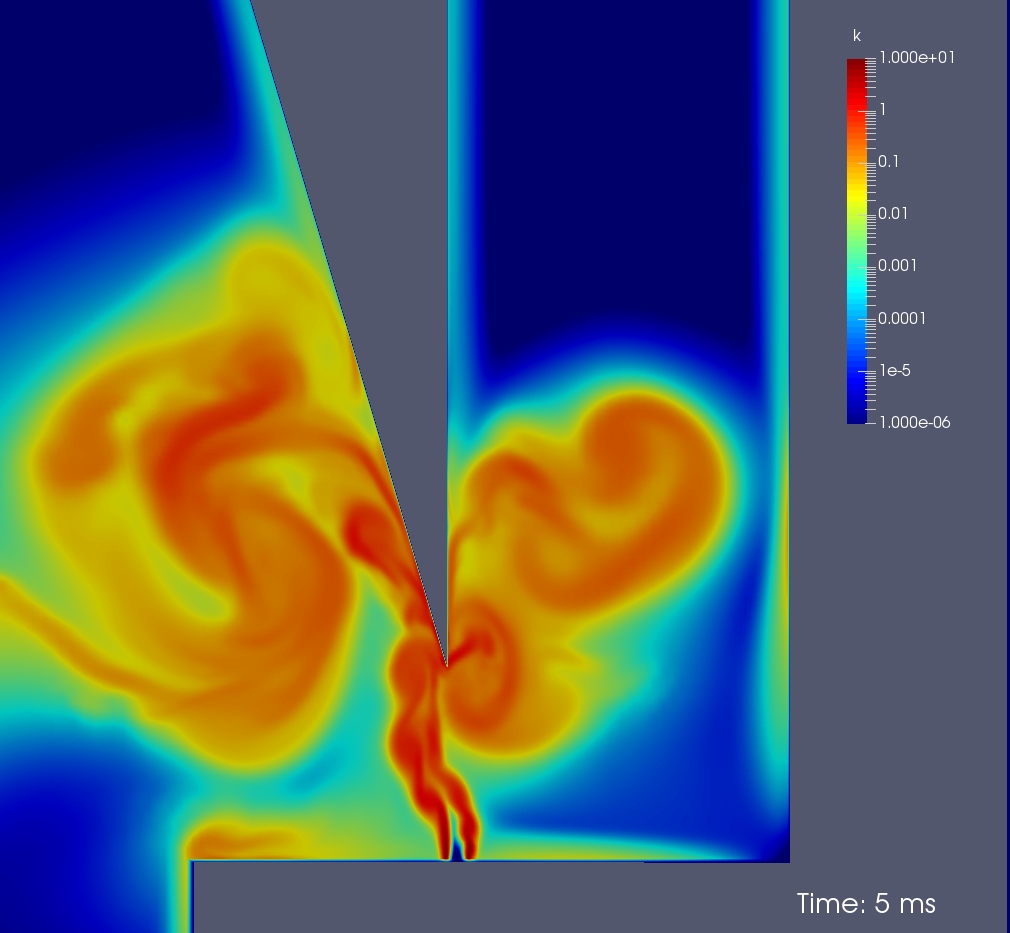}
\label{fig:k_log_01}
}
\subfigure[]
{
\includegraphics[draft=false,width=0.31\textwidth]{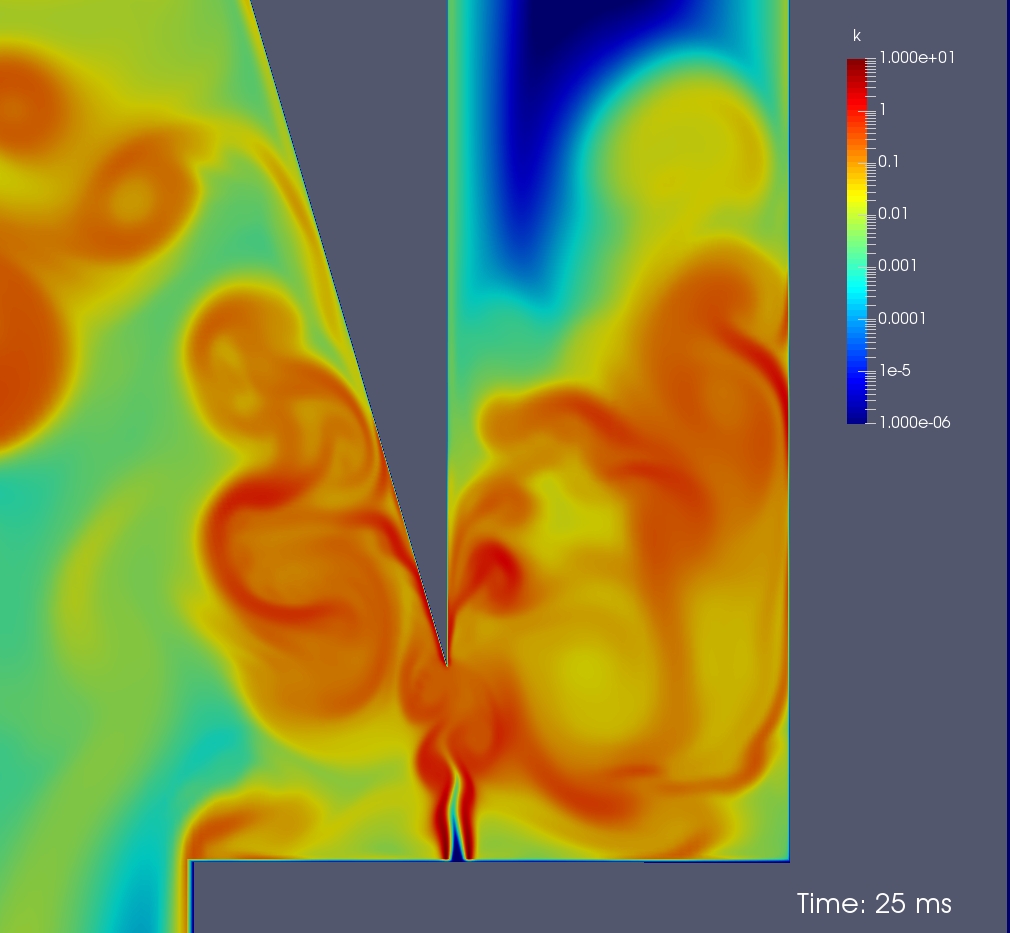}
 \label{fig:k_log_02}
}

\subfigure[]
{
\includegraphics[draft=false,width=0.31\textwidth]{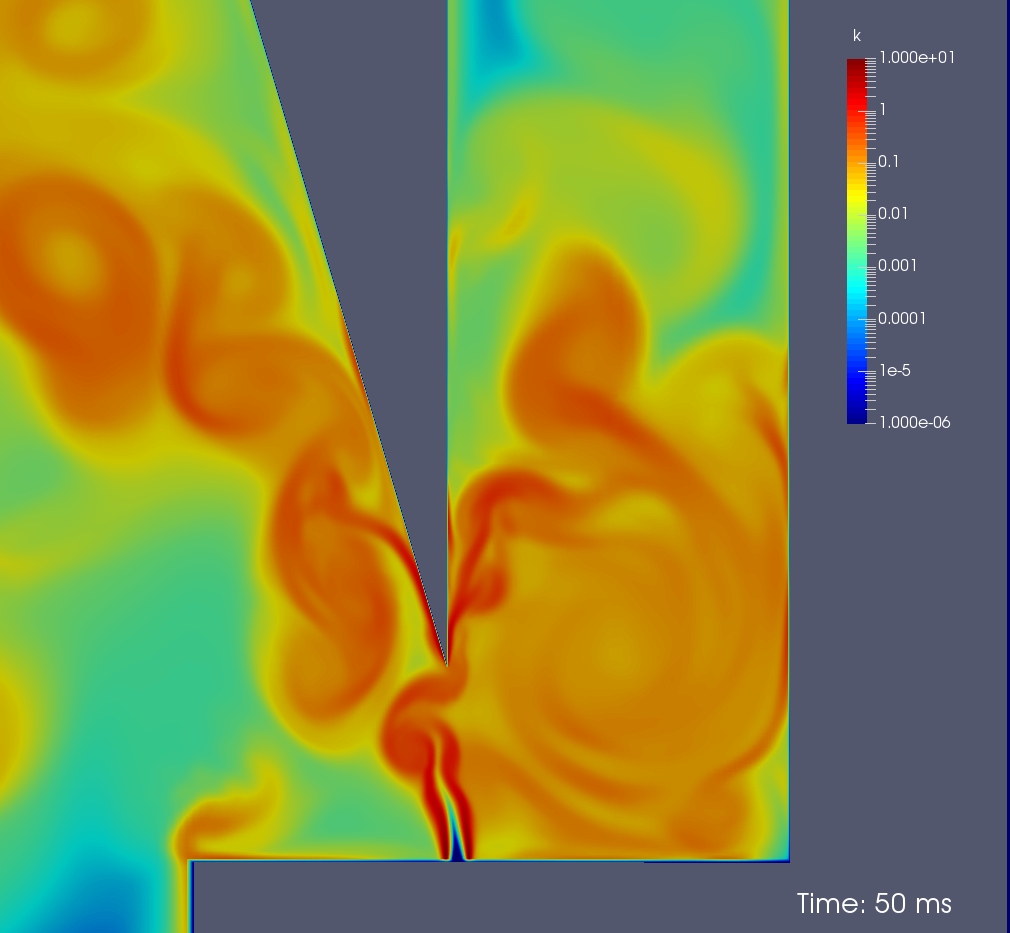}
\label{fig:k_log_03}
}
\subfigure[]
{
\includegraphics[draft=false,width=0.31\textwidth]{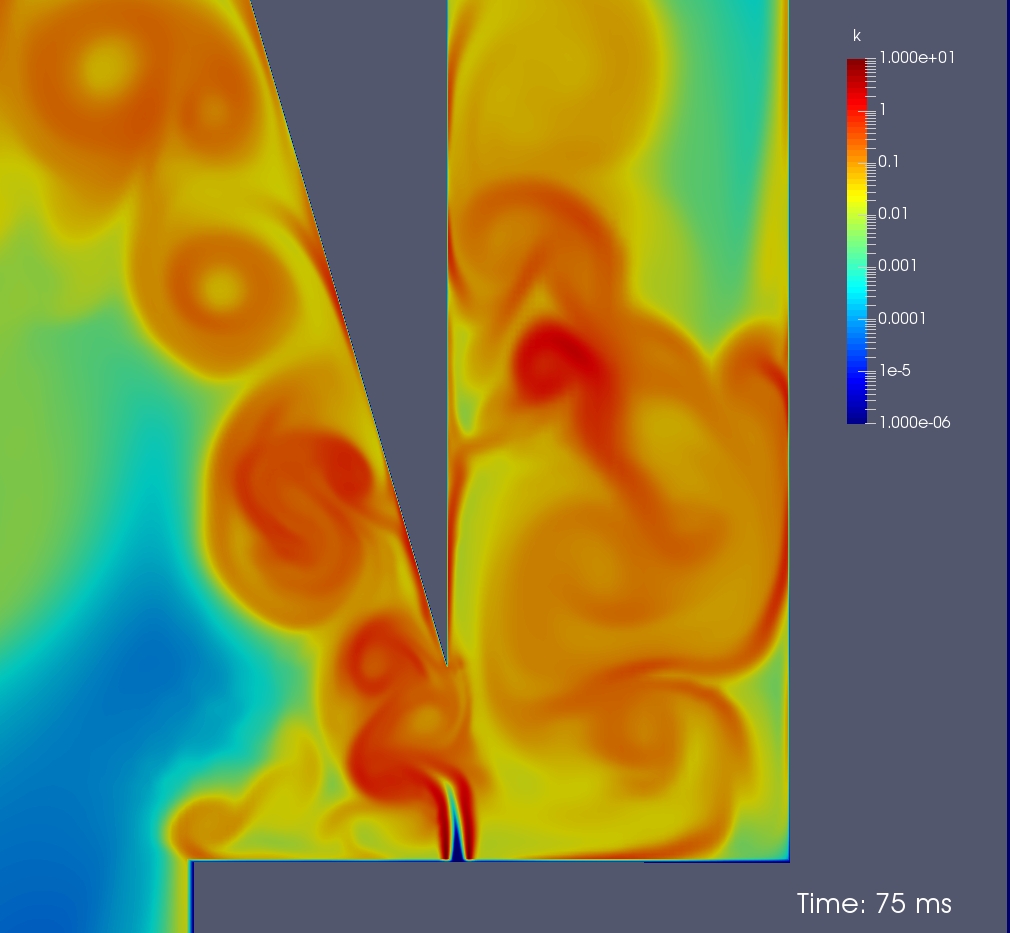}
 \label{fig:k_log_04}
}

\subfigure[]
{
\includegraphics[draft=false,width=0.31\textwidth]{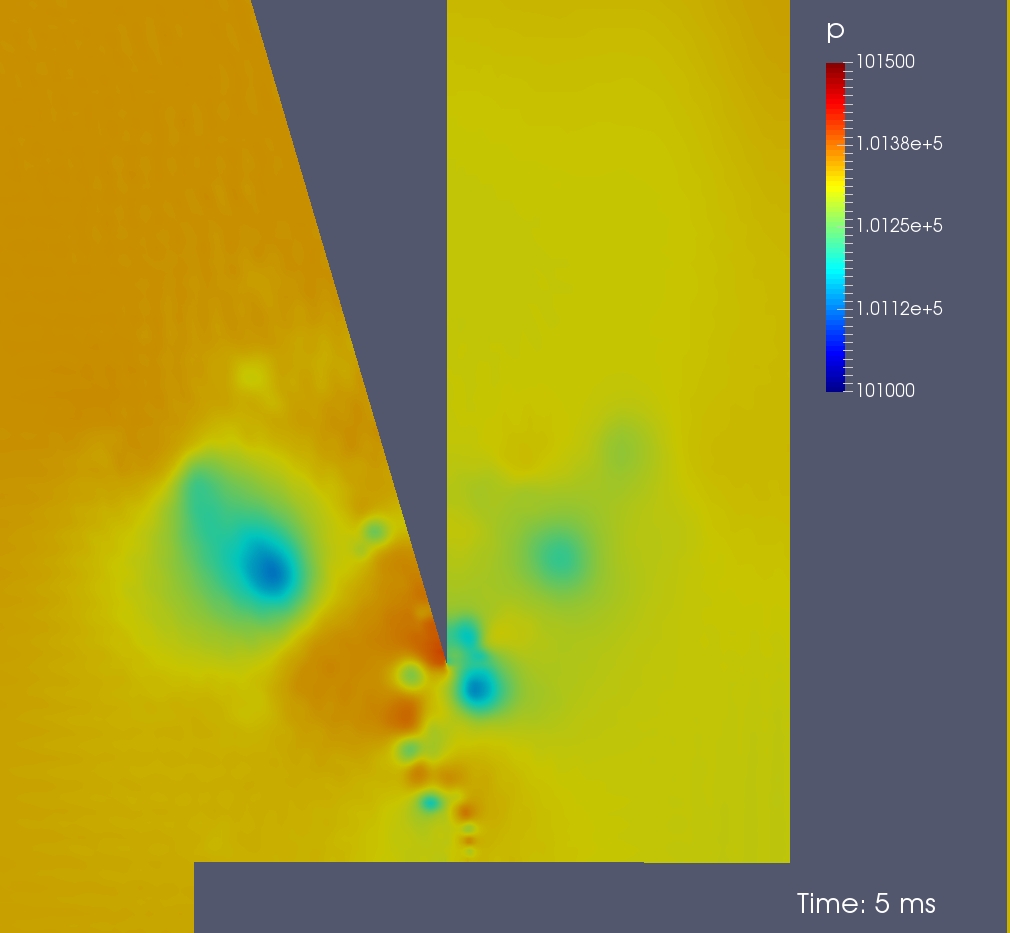}
\label{fig:detail_p_res_5ms}
}
\subfigure[]
{
\includegraphics[draft=false,width=0.31\textwidth]{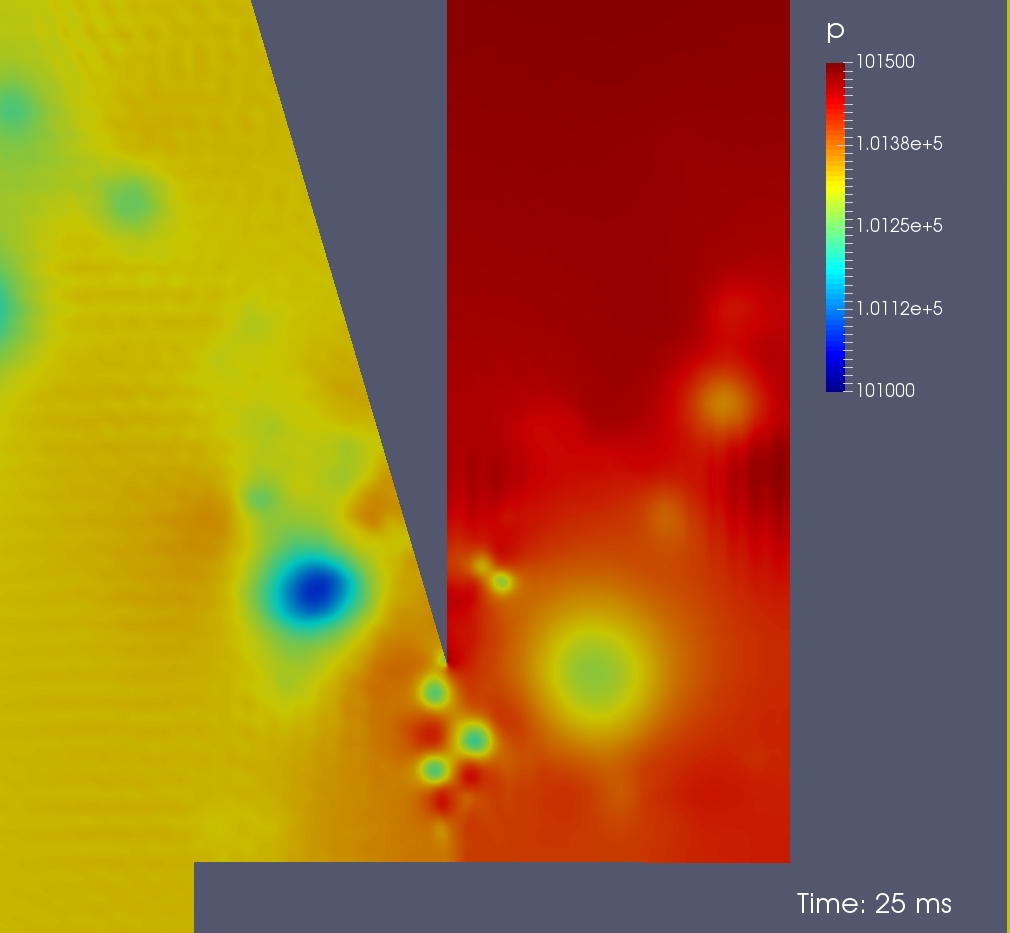}
 \label{fig:detail_p_res_25ms}
}

\subfigure[]
{
\includegraphics[draft=false,width=0.31\textwidth]{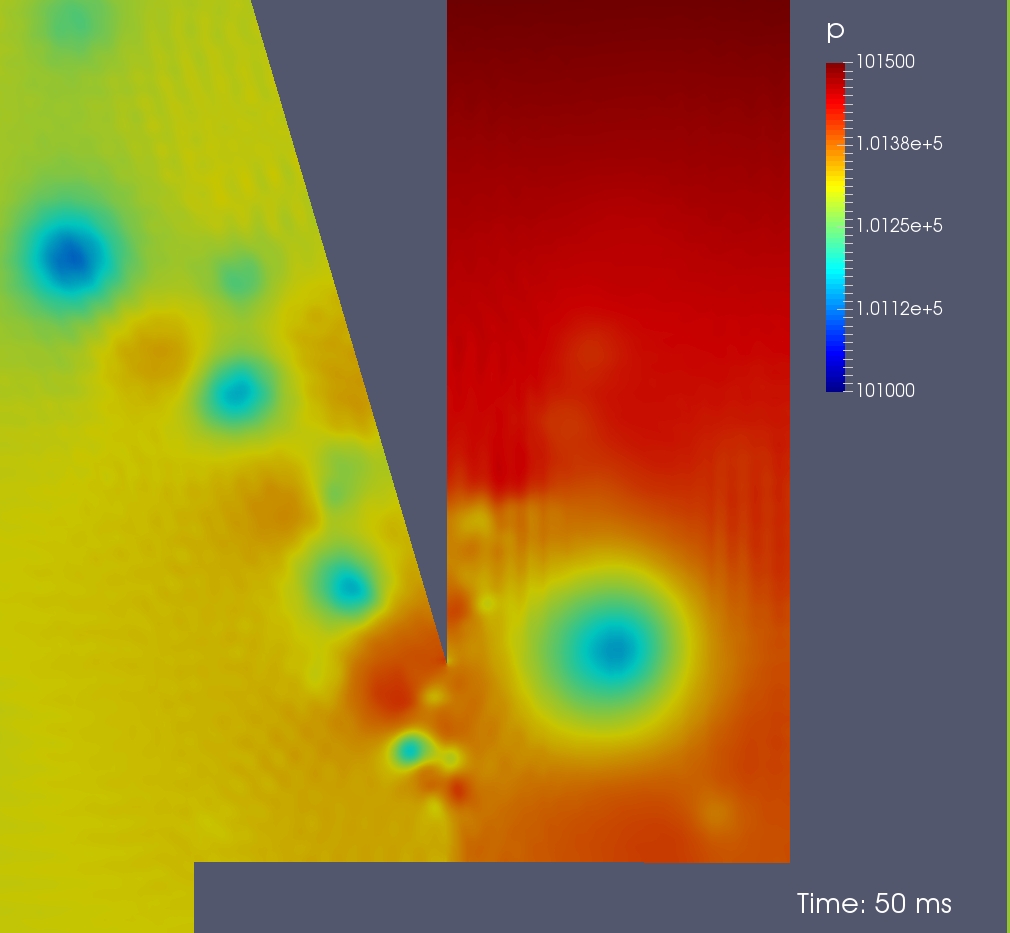}
\label{fig:detail_p_res_50ms}
}
\subfigure[]
{
\includegraphics[draft=false,width=0.31\textwidth]{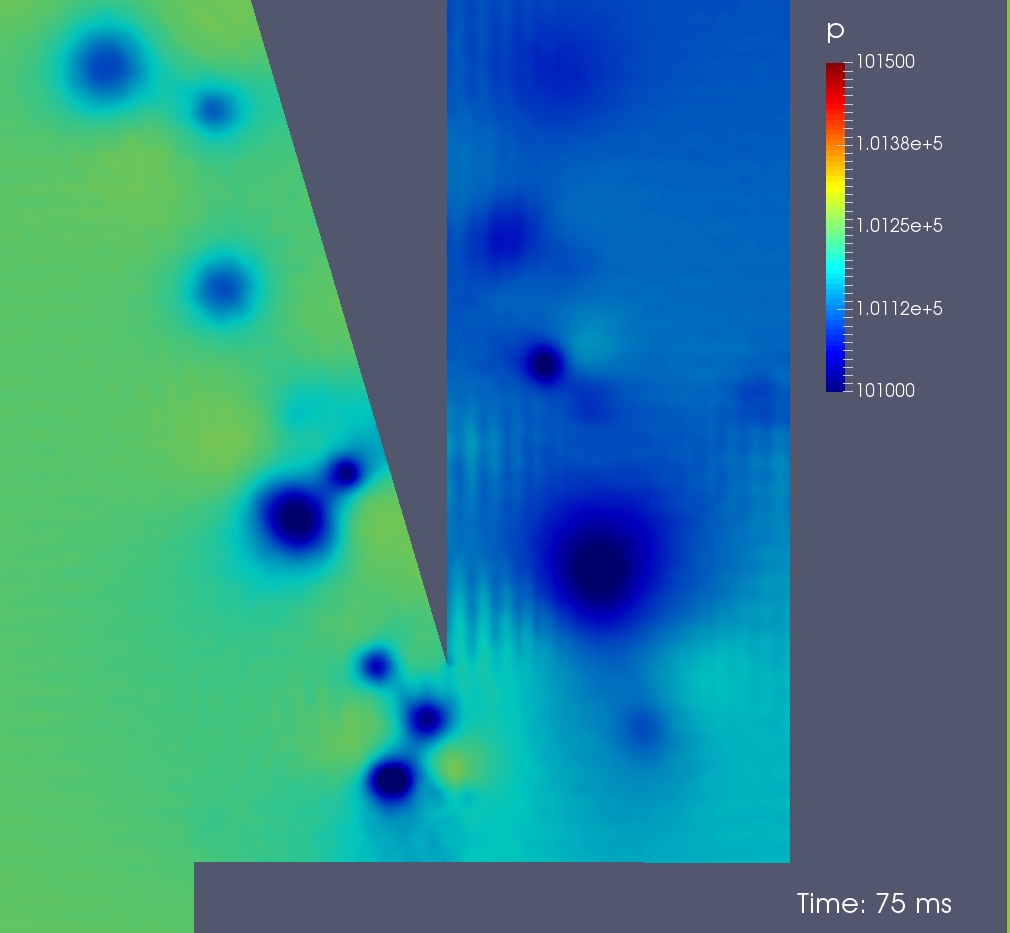}
 \label{fig:detail_p_res_75ms}
}

\caption{(a)--(d) Sequence of $t_s=5\usk\milli\second, 25\usk\milli\second, 50\usk\milli\second$ and $75\usk\milli\second$ of a numerical simulation of an operating stopped wooden organ pipe. Shown is the mouth region where the jet acts. Color-coded (log-scaled) is the turbulent kinetic energy $k$ in the range of $10^{-6}\usk\metre^2\per\second^2-10\usk\metre^2\per\second^2$. One sees the shear layers of the jet, a large vortex in the lower resonator region as well as a von K\'{a}rm\'{a}n-like vortex street peeling off outside the upper labium.  (e)--(h): The corresponding pressure $p$ is color-coded in the range of $1010\usk\hecto\pascal-1015\usk\hecto\pascal$. Clearly seen is are alternating pressure spots in the jet and a locally stable low pressure spot in the lower part of the resonator formed by the core of the (in this graphical representation clockwise) rotating vortex, referred to as the primary vortex. Outward of the upper labium a chain of alternating pressure spots form the von K\'{a}rm\'{a}n-like vortex street.}
\label{fig:k_log_sec_5-85ms}
\end{figure*}

\begin{figure*} [htb]
\subfigure[]
{
\includegraphics[draft=false,width=0.25\textwidth]{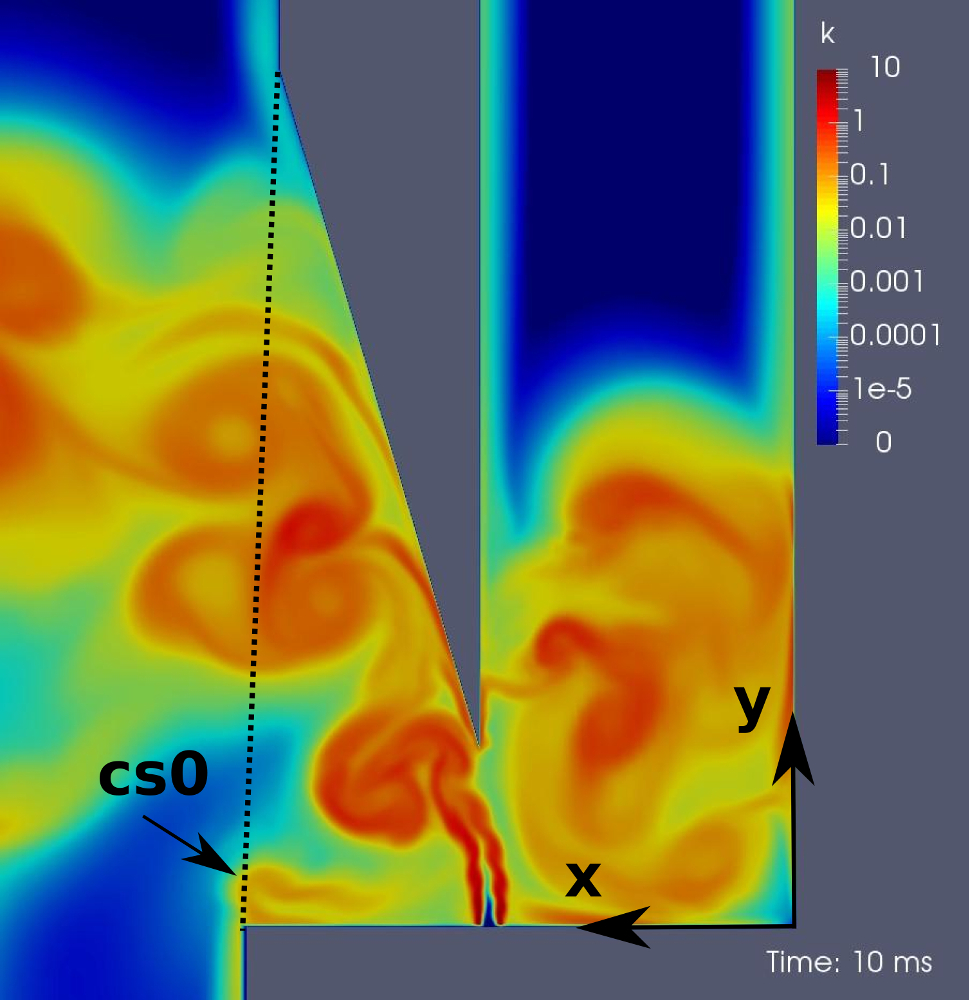}
\label{fig:numSim_03}
}
\subfigure[]
{
\includegraphics[draft=false,width=0.700\textwidth]{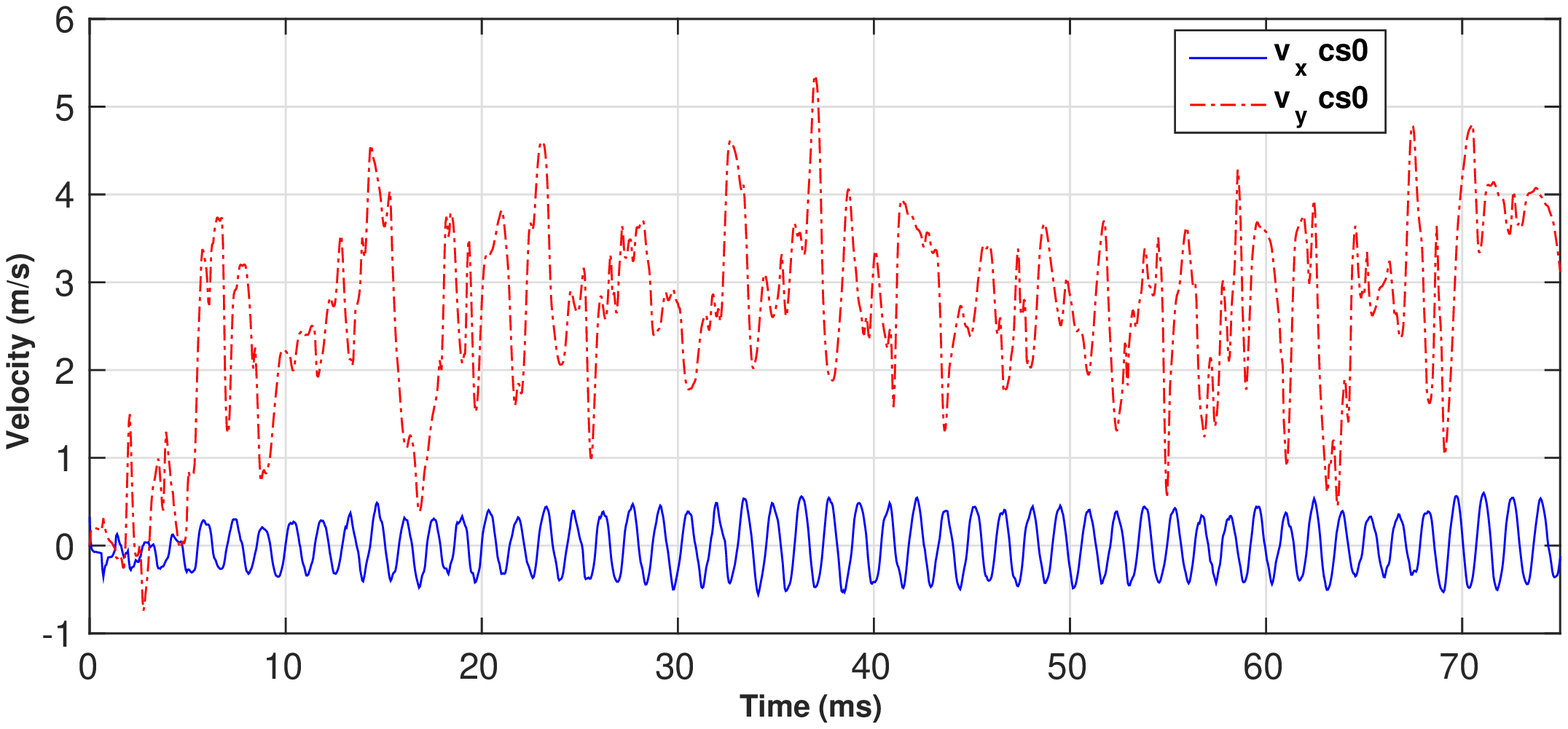}
\label{fig:signals_vx_vy}
}
\caption{(a) The mouth region of the simulated organ pipe. The snapshot at $t_s=10\usk\milli\second$ shows the jet displaced by an outgoing sound wave. Color-coded (log-scaled) is the turbulent kinetic energy $k$. From the cross-section $cs0$ (dashed line) data sets of the velocity components $v_x$ and $v_y$ and $p$ are sampled. (b) Time series of the coarse-grained velocity components $v_x$ and $v_y$, sampled at $cs0$. }
\end{figure*}

Sequences of the dynamics in the mouth region at time $t_s=5\usk\milli\second, 25\usk\milli\second, 50\usk\milli\second$ and $75\usk\milli\second$ are shown in Figs.\,\ref{fig:k_log_01}--\ref{fig:detail_p_res_75ms}. The turbulent kinetic energy $k$, color-coded (log-scaled) in a range of $10^{-6}-10\usk\metre^2\per\second^2$ , cf. Figs.\,\ref{fig:k_log_01}--\ref{fig:k_log_04}, is compared with the corresponding pressure $p$, cf. Figs.\,\ref{fig:detail_p_res_5ms}--\ref{fig:detail_p_res_75ms}. 
One can observe the two shear layers of the oscillating jet characterized by high amounts of turbulent kinetic energy. Responsible for that are the velocity profiles of the shear layers that have each an inflection point spanwise leading to the production of maximal spanwise vorticity, both typical for Kelvin-Helmholtz instabilities\cite{bailly2015turbulence}. As a result alternating pressure spots occur in the jet, increasing in diameter with respect to the propagation length. In addition further turbulent coherent structures can be observed, in particular vortices inside the lower part of the resonator. Localized in the lower part of the resonator a large rotating vortex occurs, hereafter referred to as the primary vortex. Its rotation, which is clockwise in the shown graphical representation, is driven by the periodically entering mass flux of the flow field of the oscillating jet.  The vortex core constitutes a locally stable low pressure spot in the lower resonator region, cf. Figs.\,\ref{fig:detail_p_res_5ms}--\ref{fig:detail_p_res_75ms}. The pressure gradient between the vortex core and the mouth promotes the jet's relaxation, in particular its periodical re-entering into the resonator. Along the outside of the upper labium a von K\'{a}rm\'{a}n-like vortex street peels off. It is powered by the periodical displacements of the jet and the resulting mass flux in outward direction. The energy of the jet's oscillations is supplied by the pressure difference at the windway, such that the oscillator carries its own power supply. 
To characterize the velocity field close to the jet, sample sets of various points and cross-sections were extracted from the data. For the following analysis the cross-section $cs0$ spanning the mouth region of the organ pipe is of particular interest, cf. Fig.\,\ref{fig:numSim_03}.

\section{Analysis}
\label{sec:analysis}
The visualizations show the manifold dynamics in the mouth region. The oscillating air jet disturbed by the upper labium, the rotating primary vortex in the lower resonator region and the von K\'{a}rm\'{a}n-like vortex street peeling off outside the upper labium, all of them fluid mechanical turbulent coherent structures that interact with each other in a possibly nonlinear way. To get a simpler picture one has to reduce the inherent complexity. Therefore methods of coarse-graining are utilized. The data of relevant physical quantities, such as the velocity components $v_x$, $v_y$ and the pressure $p$ were extracted at the cross-section $cs0$. It is located approximately orthogonal to the direction in which the organ pipe radiates sound waves, cf. Fig.\,\ref{fig:numSim_03}.

The spatial expansion of $cs0$ is $27.2\usk\milli\metre$. With $272$ sample points per time step the sample sets of $cs0$ have a spatial resolution of $0.1\usk\milli\metre$. A total of $1500$ time steps with temporal resolution of $5\cdot 10^{-5}\usk\second$ corresponding to $75\usk\milli\second$ were extracted. The sample sets were coarse-grained by spatial averaging (spatial integration over $cs0$ and dividing by the number of sample points) for every time step and were transferred into time series afterwards. 

In Fig.\,\ref{fig:signals_vx_vy} the time series of $v_x$ and $v_y$ are depicted for comparison. Note that $v_x$ has a strong periodicity, corresponding to a frequency of $700\usk\hertz$. The mean amplitude of $v_x$ is ca. $\langle |v_x| \rangle\simeq 0.25\usk\metre\per\second$. The velocity component $v_y$ is much more irregular and with mean amplitudes of ca. $\langle |v_y| \rangle\simeq 1.25\usk\metre\per\second$ about five times higher than those of $v_x$.

\begin{figure*} [!htb]
\subfigure[]
{
\includegraphics[draft=false,width=0.6\textwidth]{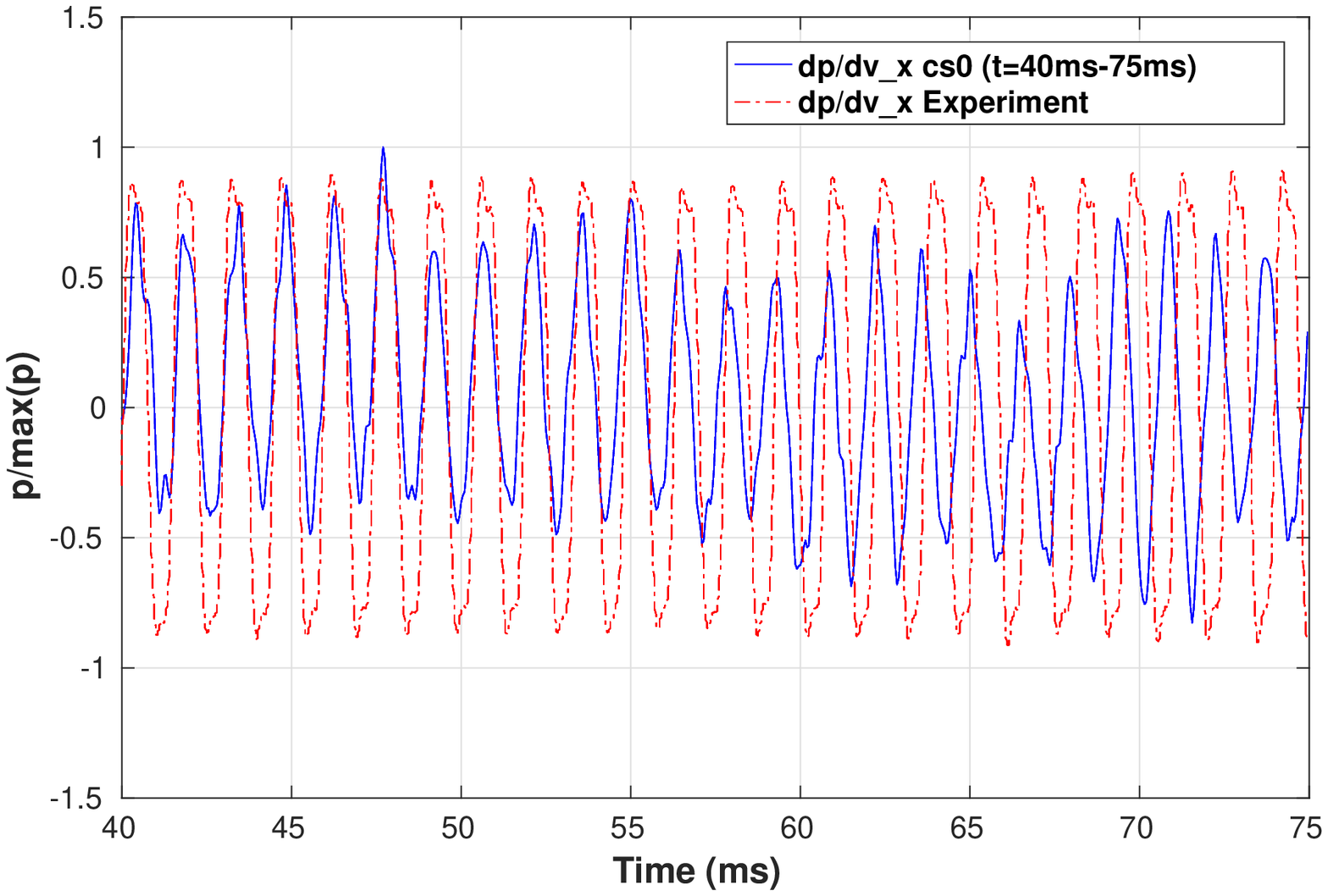}
\label{fig:p_cs0_experiment}
}
\subfigure[]
{
\includegraphics[draft=false,width=0.38\textwidth]{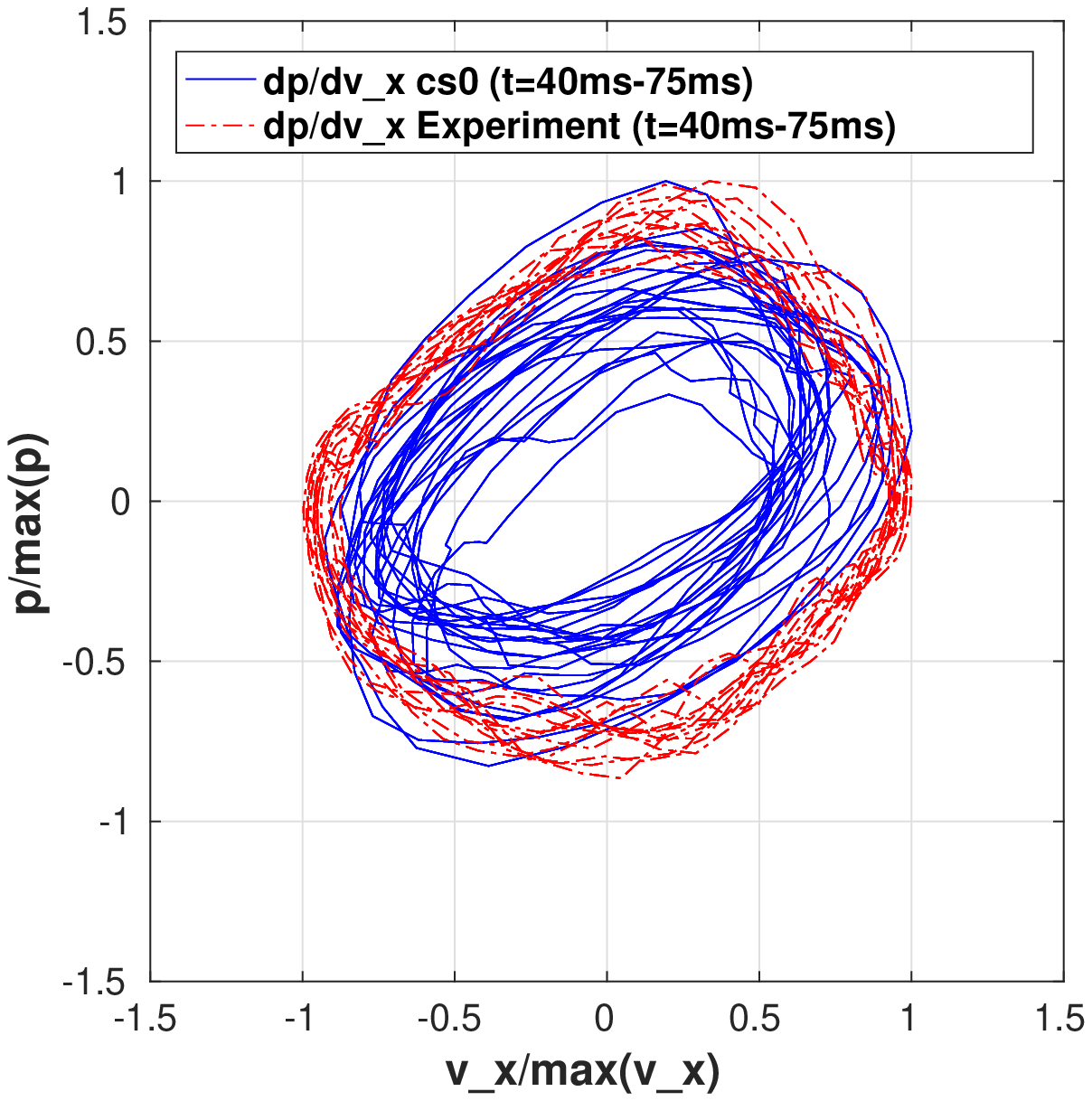}
\label{fig:phaseportrait_p_vs_v}
}

\caption{(a) The coarse-grained signal of $p$ at $cs0$ (blue line) compared with experimental data (dashed red line). (b) The coarse-grained signals of sound pressure $p$ and particle velocity $v_x$ at cross-section $cs0$ shown in the phase space representation. The trajectories shape deformed circles which indicate nonlinearities in the system.}
\label{fig:Signals_vx_vy}
\end{figure*}

The normalized time series of the pressure recorded in the experiment is compared with the coarse-grained and normalized pressure data of the cross-section $cs0$ of the numerical simulation, cf. Fig.~\ref{fig:p_cs0_experiment}.

\begin{figure*} [!htb]

\includegraphics[draft=false,width=1\textwidth]{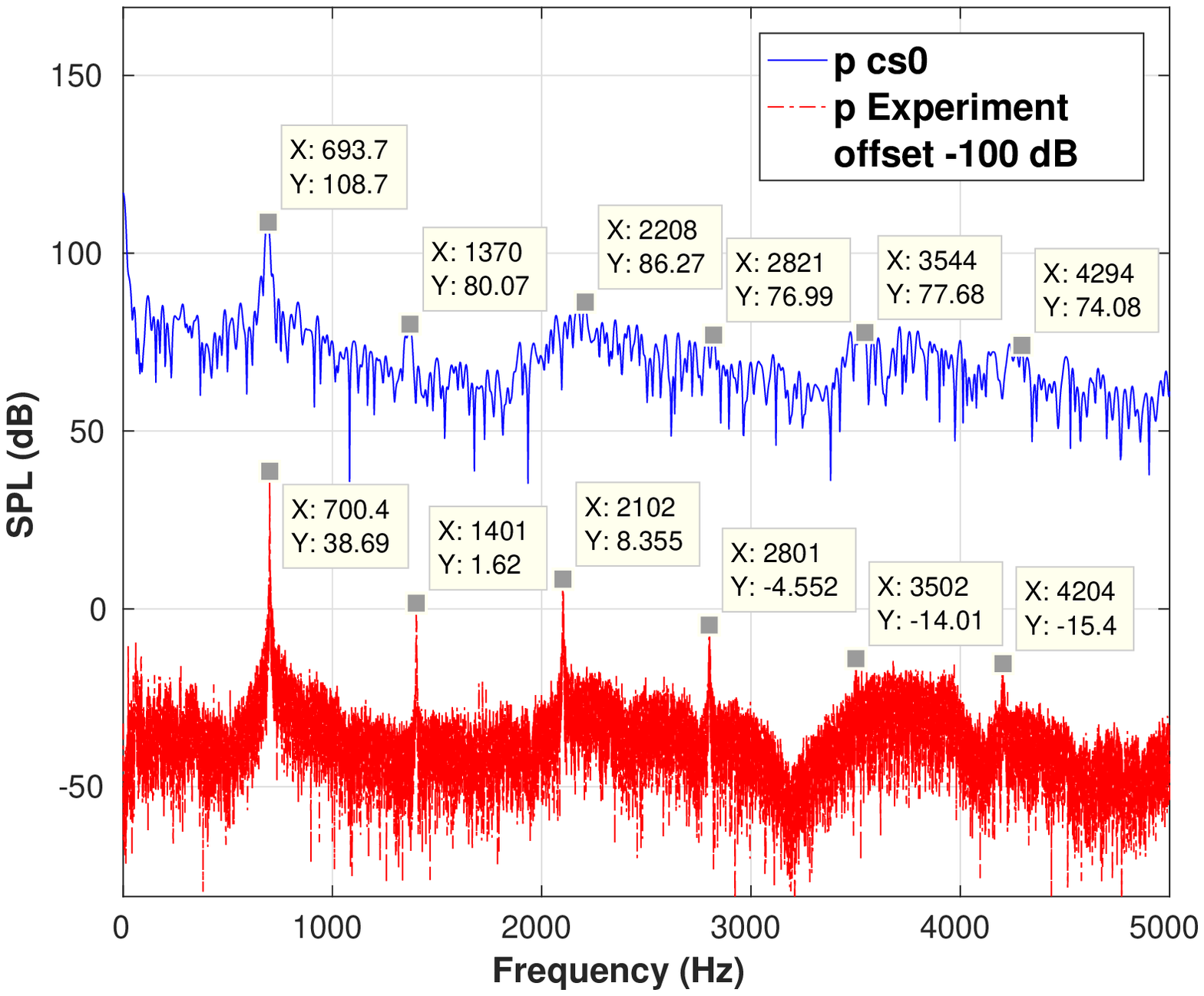}

\caption{SPL-spectrum of the coarse-grained signal of $p$ at $cs0$ (blue line) compared with experimental data (dashed red line). Labeled are the peaks of the $1^\text{th}$, $3^\text{rd}$, $5^\text{th}$ harmonics as well as the first three even harmonics, $2^{nd}, 4^{th}$ and $6^{th}$. The peaks calculated of the data sets of the numerical simulations are in very good accordance with the experimental results.}
\label{fig:spl_p_cs0}
\end{figure*}

Fig.\,\ref{fig:phaseportrait_p_vs_v} depicts the phase space representation of the parameter ($v_x$, $p$).  One observes deformed imperfect limit cycles that indicate nonlinearities in the respective extended dynamical systems. In agreement with the theory\cite{Pikovsky-Rosenblum-Kurths-01} as well as previous experimental studies\cite{Abel-Ahnert-Bergweiler-09a} the observations are typical for self-sustained oscillator systems which have intrinsic nonlinearities, often in the damping terms\cite{Abel-Ahnert-Bergweiler-09a}.

In Fig.\,\ref{fig:spl_p_cs0} the SPL-spectrum of the coarse-grained pressure signal at $cs0$ (blue line) is compared with those of experimental data (dashed red line). The peaks of the $1^\text{th}$, $3^\text{rd}$ and $5^\text{th}$ and the $2^\text{nd}$, $4^\text{th}$ and $6^\text{th}$ harmonics are labeled. Compared with the measurements of a real organ pipe the results of the numerical simulation are in good accordance.

For instance, the frequency of the third harmonic deviates by only $5\text{\%}$ from the actual measurements. Also the even harmonics, $2^{nd}$, $4^{th}$ and $6^{th}$ with lower amplitudes, occurring in real stopped wooden organ pipes, are reproduced very well by the numerical simulations. The utilized averaged values for the cross-section and the short simulation time lead to peaks that are not as prominent as in the experimental data. The data of the experiment are received by a microphone at distance $0.5\,m$ apart from the mouth region of the organ pipe. This explains the differences in amplitude between numerical simulation and experiment. Also the fact that a 2D numerical simulation
cannot map all the details of a real 3D case has to take into account at evaluating the results of the study presented here. Another source of error are numerical schemes for discretization as well as the turbulence model that models the real situation. 

\section{Results and Discussion}
\label{sec:results_and_discussion}

The main result of the study is that the flow field and the sound field in the mouth region are separable in the following respect: the velocity component $v_x$, transverse (spanwise) to the jet's flow, can be identified as carrier of particle velocity (acoustical field), while the velocity component $v_y$ with its high fluctuations in amplitude mainly consists of flow velocity (wind field), cf. Fig.~\ref{fig:signals_vx_vy}.

The aeroacoustical sensitivity of the jet, namely its periodical displacements spanwise to the jet's flow, caused by back-propagating sound waves arriving from the resonator with a fixed length, defines the pitch of a single, externally non-driven, operating organ pipe, cf. Fig.\,\ref{fig:Signals_vx_vy} and Fig.\,\ref{fig:spl_p_cs0}. This result confirms the hypothesis given in section \ref{sec:implementation} that the vortex shedding frequency corresponds to the fundamental frequency of the organ pipe. 

The pitch can change despite the fixed length of the resonator tube if the organ pipe is driven aeroacustically by an external sound source\cite{Abel-Ahnert-Bergweiler-09a}. Under special circumstances, namely if the driving frequency and the fundamental frequency of the driven organ pipe differ slightly, the coupled system can synchronize. This effect is nonlinearly dependent on the separation distance of the organ pipe and the external sound source, as was demonstrated recently\cite{fischer_synchronization_2016}.  

The mutual interaction of the internal system jet-resonator of an organ pipe is building up in the transient process. In the lower resonator region a rotating primary vortex occurs. As shown in Figs.\,\ref{fig:k_log_01}--\ref{fig:detail_p_res_75ms}, the vortex core establishes a locally stable low pressure spot so that the pressure gradient between the vortex core and the mouth promotes the jet's relaxation and therefore its re-entering into the resonator. A detailed analysis of the interplay of the turbulent coherent structures jet, primary vortex and the von K\'{a}rm\'{a}n-like vortex street,  with each other and with the passive elements, namely the resonator, the geometry of the windway, nicks or notches and the upper labium, whose position defines the cut-up's width and height and therefore the propagation length of the jet in the mouth region, is subject of current research. The presented results of the numerical simulations show an excellent robustness. 

In combination with the introduced methods, these results presented can be used to verify coarse-grained, or lumped model approaches\cite{Abel-Ahnert-Bergweiler-09a,fischer2014nichtlineare} describing the dynamics of organ pipes and their mutual aeroacoustical coupling\cite{fischer_synchronization_2016}. 

Aeroacoustical coupling of organ pipes have high practical relevance for organ builders. Organ pipes of the same stop and with nearly identical timbre are often arranged closely on the wind-chest, where they can interact with each other via the radiated sound waves. In the worst case, special sets of intonated ranks of organ stops such as Unda Maris or Voix c\'{e}leste, which consist of slightly detuned sets of organ pipes, can synchronize to the effect that their frequencies re-adjust to a single locked frequency. This unintended side-effect is known as mode locking. 

As a conclusion it can be assumed with sufficient certainty that, for modeling aeroacoustical coupling of orgen pipes, the transverse component of the velocity field of the jets is of primary relevance because these components substantially carry the particle velocity which corresponds to the sound pressure the organ pipes radiate. Eventually, the results may contribute to improve the understanding of the basic principles of sound generation as well as synchronization phenomena of organ pipes and other aerophones.\\

\subsection*{Acknowledgement}

The authors herewith acknowledge, and express their gratitude for, inspiring discussions and many helpful remarks with A. Pikovsky, M. Rosenblum from University of Potsdam. Many thanks to Alexander Schuke Orgelbau GmbH for their active help in pipe and wind supply construction.

\small
\bibliographystyle{plain}

\begin{thebibliography}{1}
\providecommand{\natexlab}[1]{#1}
\providecommand{\url}[1]{\texttt{#1}}
\expandafter\ifx\csname urlstyle\endcsname\relax
 \providecommand{\doi}[1]{doi: #1}\else
 \providecommand{\doi}{doi: \begingroup \urlstyle{rm}\Url}\fi
 
\bibitem[Verge et al. (1994)] {Verge_1994}
M. P. Verge, B. Fabre, W. E. A. Mahu, A. Hirschberg, R. R. Van Hassel, A. P. J. Wijnands, C. J. Hogendoorn. 
\newblock \emph{Jet formation and jet velocity fluctuations in a flue organ pipe.}
\newblock \emph{ The Journal of the Acoustical Society of America, 95(2), pp. 1119-1132, (1994).}
 
 
\bibitem[Fabre and Hirschberg (1997)] {Fabre-97}
M. P. Verge, B. Fabre, A. Hirschberg, A. P. J. Wijnands. 
\newblock \emph{Sound production in recorderlike instruments. I. Dimensionless amplitude of the internal acoustic field. }
\newblock \emph{The Journal of the Acoustical Society of America, 101(5), pp. 2914-2924, (1997).} 

 
\bibitem[Morse (1968)] {morse1968theoretical}
P. M. Morse, K. U. Ingard.
\newblock \emph{Theoretical acoustics.}
\newblock \emph{Princeton University Press, pp. 1-927, (1968).}


\bibitem[Schlichting (2003)] {schlichting2003}
H. Schlichting, K. Gersten. 
\newblock \emph{Boundary-layer theory.}
\newblock \emph{Springer Science \& Business Media, pp. 1-799, (2003).} 
 
 
\bibitem[Miyamoto (2010)] {miyamoto2010applicability}
M. Miyamoto, Y. Ito, K. Takahashi, T. Takami, T. Kobayashi, A. Nishida, M. Aoyagi.
\newblock \emph{Applicability of compressible LES to reproduction of sound vibration of an air-reed instrument.}
\newblock \emph{Proceedings of the International Symposium on Musical Acoustics, Sydney and Katoomba, Australia, pp. 37--43, (2010).} 


\bibitem[Bader (2013)]{bader2013nonlinearities}
 R. Bader.
\newblock \emph{Nonlinearities and Synchronization in Musical Acoustics and Music Psychology.}
\newblock \emph{Springer Science \& Business Media, pp. 1--458, (2013).}


\bibitem[Fabre and Hirschberg(2000)]{Fabre-00}
B. Fabre and A. Hirschberg.
\newblock \emph{Physical modeling of flue instruments: A review of lumped models.}
\newblock \emph{Acustica - Acta Acustica, 86: pp. 599--610, (2000).}


\bibitem[Fabre et~al.(1996)Fabre, Hirschberg, and
  Wijnands]{Fabre-Hirschberg-96}
B. Fabre, A. Hirschberg, and A. P. J. Wijnands.
\newblock \emph{Vortex shedding in steady oscillation of a flue organ pipe.}
\newblock \emph{Acustica~-~Acta Acustica, 82, pp. 863--877, (1996).}


\bibitem[Howe (1975)]{Howe-75}
M.~S. Howe.
\newblock \emph{Contribution to the theory of aerodynamic sound, with application to
  excess jet noise and theory of the flute.}
\newblock \emph{Journal of Fluid Mechanics, 71, pp. 625--673, (1975).}


\bibitem [Fischer (2014)]{fischer2014nichtlineare}
 J. L. Fischer.
\newblock \emph{Nichtlineare Kopplungsmechanismen akustischer Oszillatoren am Beispiel der Synchronisation von Orgelpfeifen, (Nonlinear Coupling Mechanisms of Acoustical Oscillators by the Example of the Synchronization of Organ Pipes). PhD-Thesis.}
\newblock \emph{University Library Potsdam, pp. 1--255, (2014).}


\bibitem[OpenFoam (2017)] {openfoam_guide}
OpenFOAM$^\circledR$ - The Open Source Computational Fluid Dynamics (CFD) Toolbox
Organization - OpenCFD Limited.
\newblock URL \url{http://www.openfoam.com/} \emph{, (2017).}


\bibitem[Bailly (2015)]{bailly2015turbulence}
Ch. Bailly, G. Comte-Bellot.
\newblock \emph{Turbulence},
\newblock \emph{Springer Science \& Business Media, pp. 21--27, (2015).}


\bibitem[Kolmogorov (1941)] {kolmogorov1941a}
A. N. Kolmogorov. 
\newblock \emph{The local structure of turbulence in incompressible viscous fluid for very large Reynolds numbers.}
\newblock \emph{Dokl. Akad. Wiss. USSR, Bd. 30, pp. 301--305, (1941).}


\bibitem[Schuke (2016)]{Schuke}
{A}lexander {S}chuke Orgelbau {G}mb{H}.
\newblock URL \url{http://www.schuke.de/} \emph{,(2017).}


\bibitem[Pikovsky et~al.(2001)Pikovsky, Rosenblum, and
  Kurths]{Pikovsky-Rosenblum-Kurths-01}
A. Pikovsky, M. Rosenblum, and J.~Kurths.
\newblock \emph{Synchronization---A Universal Concept in Nonlinear Science.}
\newblock \emph{Springer, Berlin, pp. 1--411, (2001).}


\bibitem[Abel et~al.(2009)Abel, Ahnert, and Bergweiler]{Abel-Ahnert-Bergweiler-09a}
M.~Abel, K.~Ahnert, and S.~Bergweiler.
\newblock \emph{Synchronization of sound sources.}
\newblock \emph{Physical Review Letters, 103\penalty0 (114301), (2009).}


\bibitem[Fischer et~al.(2017)]{fischer_synchronization_2016}
J. L. Fischer, R. Bader, M. Abel. 
\newblock \emph{Aeroacoustical coupling and synchronization of organ pipes.}
\newblock \emph{The Journal of the Acoustical Society of America, 140(4), pp. 2344-2351, (2016).}

\end{thebibliography}


\end{document}